\documentclass[prb,twocolumn,superscriptaddress,preprintnumbers,amsmath,amssymb]{revtex4}
\usepackage{graphicx}
\usepackage{amsmath}
\usepackage{amssymb}
\usepackage{epsfig}
\usepackage{wasysym}
\usepackage{bbm}
\usepackage{multirow}
\renewcommand{\narrowtext}{\begin{multicols}{2} \global\columnwidth20.5pc}
\renewcommand{\v}[1]{{\bf #1}}

\def\be{\begin{eqnarray}}
\def\ee{\end{eqnarray}}
\newcommand{\nn}{\nonumber\\}
\newcommand{\Eq}[1]{Eq.~(\ref{#1})}
\newcommand{\Fig}[1]{Fig.~(\ref{#1})}

\newcommand{\nd}{{\vphantom{\dagger}}}

\begin{document}

\title{Antiferromagnetically Driven Electronic Correlation in Iron Pnictides and Cuprates}

\author{Hui Zhai}
\affiliation{
Department of Physics,University of California at Berkeley,
Berkeley, CA 94720, USA}
\affiliation{Materials Sciences Division,
Lawrence Berkeley National Laboratory, Berkeley, CA 94720, USA}
\author{Fa Wang}
\affiliation{
Department of Physics,University of California at Berkeley,
Berkeley, CA 94720, USA}
\author{Dung-Hai Lee}
\affiliation{
Department of Physics,University of California at Berkeley,
Berkeley, CA 94720, USA}
\affiliation{Materials Sciences Division,
Lawrence Berkeley National Laboratory, Berkeley, CA 94720, USA}

\begin{abstract}

The iron pnictides and the cuprates represent two families of materials, where strong antiferromagnetic correlation drives  three other distinct ordering tendencies: (1) superconducting pairing, (2) Fermi surface distortion, and (3) orbital current order. We propose that (1)-(3) and the antiferromagnetic correlation are the hallmarks of a class of strongly correlated materials to which the cuprates and pnictides belong. In this paper we present the results of the functional renormalization group studies to support the above claim. In addition, we show that as a function of the interlayer hopping parameter, the double layer Hubbard model nicely interpolates between the cuprate and the iron pnictide physics. Finally, as a check, we will present the renormalization group study of a ladder version of the iron pnictide, and compare the results to those of the two-dimensional model.

\end{abstract}

\date{\today}
\maketitle

\tableofcontents

\section{Introduction}

The discovery of high-temperature superconductivity in the iron pnictides adds a new class of compounds to the high-T$_c$ family \cite{iron,NLWang,haihu,122}. In many regards, the pnictides are very similar to the cuprates. For instance, the undoped compound shows antiferromagnetic (AFM) long-range order and becomes superconducting (SC) upon doping, and the doping-temperature phase diagrams is similar to the cuprates.  On the other hand there are important differences. For example, the undoped cuprates are Mott insulators while the undoped pnictides are semi-metals.

In addition to the cuprates and the pnictides, there are other materials that become superconducting upon exiting the antiferromagnetic phase. Examples include the heavy fermion compounds \cite{heavyfermion} and the organic compounds \cite{organic} (in these systems pressure  replaces doping as a tuning parameter).

An important question concerning these materials is ``does antiferromagnetic correlation have anything to do with the superconducting pairing''. In this paper, we try to answer this question by performing functional renormalization group (FRG) calculations. We shall focus on the iron pnictides and the {\it overdoped} cuprates. The reason for focusing on the overdoped cuprates is because we believe FRG has a better chance to succeed. As for the iron pnictides, the general belief is that its electronic correlation is weaker. We believe this makes FRG suited for the entire doping range.

Although the pnictides is not as strongly correlated as the cuprates, in some sense it is more difficult. In the cuprates there is a  charge gap, below which superexchange generates the  AFM interaction. With the AFM interaction, superconducting pairing appears in mean-field theory \cite{kotliar}.   For the iron pnictides there is no charge gap, and the bare (purely repulsive) Hamiltonian does not show any mean-field SC instability. In this case we need a renormalization group procedure (in contrast to the one-shot second order perturbation procedure for the cuprates) to generate the effective Hamiltonian. Of course, the cuprate problem is difficult in another aspect, namely, at low energies there is the no-double occupancy constraint.

For the cuprates it is well recognized that it is the underdoped limit which exhibits the most unconventional behaviors. Examples include pseudaogap and the dichotomy between the nodal and antinodal electronic excitations \cite{XJZhou}. In that limit the system also shows the propensity toward charge/spin density wave order\cite{DDW,patrick,Lee-Wen}. In contrast the overdoped limit appears more conventional. The normal state looks like an ordinary metal, and the  SC state appears to be a garden variety of d-wave superconductor. Interestingly for the LSCO system it is shown that even the overdoped samples show substantial AFM correlation \cite{Birgeneau}. In the literature, the proximity to the Mott insulating limit is widely regarded as why cuprates show high T$_c$. In our opinion the iron pnictides cast some doubt on this point of view.

Now let us narrow our focus to the recent theoretical developments concerning the pnictides.
From a theoretical standpoint what makes the pnictides more complex is the fact that there are five relevant bands \cite{kuroki,Mazin,LDA}. They give rise to multiple Fermi surfaces (FS). It is conjectured early on that the Umklapp scattering between the electron and the hole FS is responsible for the SC pairing, and because the sign of the Umklapp scattering is positive, the gap function takes opposite sign on the electron and hole pockets\cite{kuroki,Mazin}. This conclusion is independently reached by a number of other groups\cite{Fa,mechanism,Jianxin,Chubukov} by performing calculations with different degree of approximation. Among them the most unbiased treatment is the FRG of Ref.~\cite{Fa}. This pairing symmetry and mechanism is supported by recent ARPES experiments \cite{no-node,DingHong}. In addition, the lack of the Hebel-Slichiter peak in NMR experiments \cite{HS} and the presence of a neutron resonance peak \cite{resonance} in the SC state are both consistent with the $s_{\pm}$ pairing symmetry \cite{scalapino}.

The above approaches start from the band semi-metal are sometimes regarded as the ``weak coupling'' approach. While this phrase is justified for some of the above approaches, it is not for the FRG. Indeed, as a function of decreasing energy cutoff various scattering amplitudes grow in strength. By the time only a thin shell around each Fermi surface is left, these scattering can get very strong. In addition, the quasiparticles contained in these thin shells are no longer bare electrons.
For example the coupling of these quasiparticles to the staggered magnetic field is enhanced compared to bare electrons.

There is also another approach which starts with a $J_1-J_2$ model which can produce the observed $(\pi,0)$ or $(0,\pi)$ AFM ordering and the associated four-fold rotation symmetry  breaking \cite{neutron,JP,cenke}. Upon doping, mean-field theory similar to those of the cuprates (except there is no occupation constraint) gives rise to the same out-of-phase SC pairing\cite{JP}.  However, since for iron pnictides there is no charge gap to justify the superexchange, the  validity of the effective model requires a justification. In addition to the above, there are many other theoretical works addressing various aspects of the pnictides \cite{theory_rest}.

For the cuprates and pnictides antiferromagnetism and superconductivity are incontrovertible. Moreover, strong experimental evidences indicate that the propensity toward unidirectional charge/spin density wave ordering (or ``stripe ordering'') exists for many underdoped (or sometimes even overdoped) cuprate families\cite{neutron_cup}. Theoretically the staggered orbital current order\cite{DDW, patrick} and the tendency toward Pomeranchuk instability (PI) \cite{Pomeranchuk}  has also been discussed for the cuprate materials. For example, the ``D-density wave'' order has been proposed to generate the anti-nodal pseudogap\cite{DDW}. Moreover, there are proposals that such type of order will be exhibited in the vortex core of the high T$_c$ superconducting state.\cite{patrick,Lee-Wen}. Concerning the PI it has been proposed that interaction which favors a d-wave FS distortion exists \cite{d-wave}. Moreover it is argued that when such distortion occur, it will lead to an electronic nematic state \cite{nematic} which is the precursor of the stripe ordering. In this paper we shall show that a similar group of ordering tendencies exist for the pnictides. We propose that the quartet of ordering tendencies including the AFM ordering, the SC pairing, the orbital current ordering, and the Pomeranchuk instability, are the hallmark of a whole class of correlated materials - materials where the short range electronic repulsive interaction induced strong AFM fluctuations dominate the low temperature properties.

The paper is organized as follows. In Sec. II, we shall start with the FRG results for the iron pnictides. The goal is to identify the interactions that grow at low energies, and to construct an effective Hamiltonian describing them. We shall argue that the AFM scattering is what's driving these strong interactions. In Sec III, we shall revisit the FRG result for the single-band Hubbard model proposed for the cuprates\cite{schulz,Honerkamp}. Same efforts will be spent to identify growing interaction at low energies.  Many results in this section have already been obtained before\cite{schulz,Honerkamp}. We present them for the purpose of making comparison with the pncitides results. It is from this comparison a coherent picture emerges. In this picture AFM scattering is the root of SC, PI, and orbital current ordering tendencies. In Sec IV, we show that as a function of the inter-layer hopping parameter ($t_z$) the double-layer Hubbard model nicely interpolates between the physics of the pnictides and cuprates. In particular, as a function of increasing $t_z$ the band structure of this model evolves from two large FS (with shape similar to that of the curpates) to two separate electron and hole pockets (similar to the pnictides). Upon doping the SC pairing undergoes a phase transition from d-wave to out-of-phase s-wave as a function of $t_z$.
In Sec V, we present the renormalization group calculation for a ladder version of the pnictide model. The purpose of this calculation is to provide a check of the two-dimensional FRG results presented in Sec II. This is motivated by the fact that in the case of the cuprates, much of the d-wave pairing physics is already captured by the two-leg ladder \cite{HHLin}. Section VI is the summary and conclusion.

\section{The FRG Results for the Iron Pnictides}

The goal of the FRG calculation is to generate the renormalized electron-electron scattering Hamiltonian at low energies:
\begin{equation}
V({\bf k_1},{\bf k_2},{\bf k_3},{\bf k_4},\Lambda)\psi^\dag_{{\bf k_1},s}\psi^\dag_{{\bf k_2},s^\prime}\psi_{{\bf k_3},s^\prime}\psi_{{\bf k_4},s}.\label{vf}
\end{equation}
Here $\Lambda$ is the running energy cutoff, and ${\bf k}=(k_x,k_y,\alpha)$ specifies the momentum $\vec{k}=(k_x,k_y)$ as well as the band index $\alpha$. We watch for those growing  $V({\bf k_1},{\bf k_2},{\bf k_3},{\bf k_4},\Lambda)$ as $\Lambda$ decreases.

\begin{figure}[tbp]
\begin{center}
\includegraphics[angle=0,scale=0.48]
{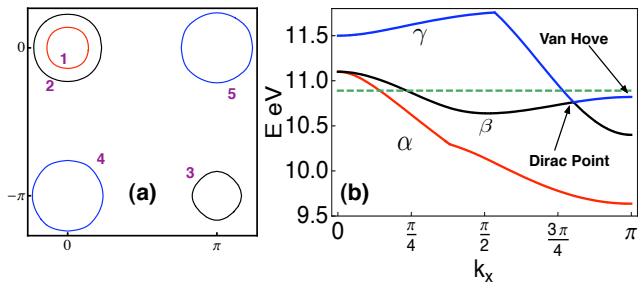}\caption{(color on-line)(a) The LDA FS of iron pnictides\cite{kuroki}. $1-3$ are hole FS and $4,5$ label the electron FS. (b) The LDA dispersion along the $(0,0)-(\pi,0)$ line. Here the red, black, and blue curves represent  the  $\alpha,\beta$ and $\gamma$ band, respectively. The dashed green line marks the Fermi level of undoped compound. The Van Hove and the Dirac band singularities discussed in the text are indicated. \label{FS}}
\end{center}
\end{figure}

For the iron pnictide there are five relevant bands \cite{kuroki,Mazin,LDA}. Two hole bands, denoted by $\alpha$ and $\beta$, give rise to two hole pockets around the $(0,0)$ point. Depending on the doping, there can be another hole pocket around $(\pi,\pi)$. (In this paper we use the so-called unfolded Brillouin zone.) The electron band, denoted by $\gamma$, gives two electron pockets: one around $(\pi,0)$ and the other around $(0,\pi)$. For the band structure given by Kuroki {\it et al.} \cite{kuroki}, the electron dispersion along $(0,0)-(\pi,0)$ and the FS are shown in Fig. \ref{FS}. To a good approximation, the FS are  nested by momenta  $(\pi,0)$ and  $(0,\pi)$ for a range of doping. A measure of the the ``failure of nesting" can be obtained as follows. Take the best nested pair of electron and hole FS, displaces the hole FS by $(\pi,0)$, record the energies of the electron band around the displaced FS, and calculate the maximum absolute value $\Delta_{\text{mis}}$ of these energies. For the band structure of Kuroki {\it et al.} \cite{kuroki}, $\Delta_{\text{mis}}\le 0.05 eV$ for $\lesssim 10\%$ hole doped or parent compound, and $\le 0.1eV$ for $\lesssim 10\%$ electron doped compounds. For $\Lambda>\Delta_{\text{mis}}$ we should regard the FS as nested. Above $\Delta_{\text{mis}}$ the generic ``on-shell'' scattering processes include Cooper scattering, forward scattering, and $(\pi,0),(0,\pi)$ spin density wave (SDW) and charge density wave (CDW) scattering.

For the processes mentioned above, we can cast the corresponding four-fermion scattering Hamiltonian into following form
\begin{equation}
\sum\limits_{{\bf k},{\bf p}}V({\bf k},{\bf p},\Lambda)\hat{O}^\dag_{{\bf k}}\hat{O}_{{\bf p}}\label{ord}
\end{equation}
where $\hat{O}_{{\bf k}}$ can be pairing operator, current density operator or spin density operator. (We shall explain this in detail later.) We then decompose $V({\bf k},{\bf p},\Lambda)$ into different eigen-modes as
\begin{equation}
V({\bf k},{\bf p},\Lambda)=\sum\limits_i w_i(\Lambda) f^{i*}({\bf k})f^i({\bf p})\label{dec}
\end{equation}
We monitor $w_i$ as a function of the energy cutoff $\Lambda$. If $w_i$ is negative and its absolute value grows as $\Lambda$ decreases, it indicates that the system has a propensity for developing the order characterized by the following order parameter
\begin{equation}
\mathcal{O}_i=\sum\limits_{{\bf k}}f^i({\bf k})\langle \hat{O}_{{\bf k}}\rangle
\end{equation}
where the form factor $f^i({\bf k})$ tells us about the symmetry of the order parameter. Since the renormalized quasiparticles (left in thin shells around the FS) are not the bare electrons, we can not unambiguously determine the real space pattern of the order. The  real space patterns that we shall present below are the simplest representation that captures the symmetry of $f^i({\bf k})$ when it is Fourier transformed to momentum space.

The Hamiltonian we shall perform FRG upon is given by
\begin{eqnarray}
&&H=\sum_{\bf{k},s}\sum_{a,b=1}^5 \psi_{a\bf{k} s}^\dagger K_{ab}(\v{k})\psi_{b\v{k} s}\nonumber\\&&+\sum_{i}\Big\{U_1\sum_{a}
 n_{i,a,\uparrow}n_{i,a,\downarrow}\nonumber\\&&+U_2\sum_{a< b}n_{i,a} n_{i,b}+
 J_H\sum_{a< b,s,s^\prime}\psi_{ia s}^\dagger \psi_{ib s'}^\dagger \psi_{ia s'}^\nd \psi_{ib s}\nonumber\\&&+J_H\sum\limits_{a<b} (\psi^\dagger_{ia\uparrow}\psi^\dagger_{ia\downarrow}\psi_{ib\downarrow}\psi_{ib\uparrow}+h.c.) \Big\}.\label{model_5band}
\end{eqnarray}
The parameters used in constructing $K_{ab}(\v k)$ can be found in Ref.~\cite{kuroki}. The results in this section is obtained by using $U_1=4 eV$ and $U_2=2 eV$ and $J_{\text{H}}=0.7eV$. The results are qualitatively the same for other choices of bare interactions. The details of our FRG calculation can be found in Ref. \cite{Fa}.

\subsection{The Spin Density Wave}

In reference to \Eq{vf}
we consider the following two types of scattering processes, one is $\vec{k}_1-\vec{Q}=\vec{k}_3=\vec{k}$ and $\vec{ k}_2-\vec{Q}=\vec{k}_4=\vec{p}$, and the other is $\vec{k}_1-\vec{Q}=\vec{k}_4=\vec{k}$ and $\vec{k}_2-\vec{Q}=\vec{ k}_3=\vec{p}$, where $\vec{Q}$ can be either $(\pi,0)$ or $(0,\pi)$. Once the momenta are fixed, the band index will be correspondingly fixed by the conservation of momentum. For example if ${\bf k}$ is on the electron (hole) band, ${\bf k+Q}$ will be on the hole (electron) band. The corresponding scattering terms are
\begin{equation}
V({\bf k+Q},{\bf p+Q},{\bf k},{\bf p},\Lambda)\psi^\dag_{{\bf k+Q},s}\psi^\dag_{{\bf p+Q},s^\prime}\psi_{{\bf k},s^\prime}\psi_{{\bf p},s},
\end{equation}
and
\begin{equation}
V({\bf k+Q},{\bf p+Q},{\bf p},{\bf k},\Lambda)\psi^\dag_{{\bf k+Q},s}\psi^\dag_{{\bf p+Q},s^\prime}\psi_{{\bf p},s^\prime}\psi_{{\bf k},s},
\end{equation}
We now cast their sum into SDW and CDW scattering as follows:
\begin{equation}
V_{\text{SDW}}({\bf k},{\bf p},\Lambda)\vec{S}_{{\bf k},Q}\vec{S}_{{\bf p},Q}+V_{\text{CDW}}({\bf k},{\bf p},\Lambda)n_{{\bf k},Q}n_{{\bf p},Q},
\end{equation}
where
\begin{equation}
\vec{S}_{{\bf k},Q}=\sum\limits_{s,s^\prime}\psi^\dag_{s,{\bf k+Q}}\vec{\sigma}_{ss^\prime}\psi_{s^\prime,{\bf k}}.
\end{equation}
and
\begin{equation}
n_{{\bf k},Q}=\sum\limits_s \psi^\dag_{s,{\bf k+Q}}\psi_{s,{\bf k}}.
\end{equation}
Using the identity
\begin{equation}
\vec{\sigma}_{ss^\prime}\vec{\sigma}_{\tau\tau^\prime}=2\delta_{s\tau^\prime}\delta_{\tau s^\prime}-\delta_{ss^\prime}\delta_{\tau\tau^\prime}
\end{equation}
one can show that
\begin{equation}
V_{\text{SDW}}({\bf k},{\bf p},\Lambda)=-\frac{1}{2}V({\bf k+Q},{\bf p+Q},{\bf k},{\bf p},\Lambda)\label{SDWvertex}
\end{equation}
and
\begin{eqnarray}
V_{\text{CDW}}({\bf k},{\bf p},\Lambda)=V({\bf k+Q},{\bf p+Q},{\bf p},{\bf k},\Lambda)\nonumber\\-\frac{1}{2}V({\bf k+Q},{\bf p+Q},{\bf k},{\bf p},\Lambda)
\label{cd}\end{eqnarray}
In the above and later section, by ``CDW'' we mean the $(\pi,0)$ or $(0,\pi)$  density wave in the spin singlet channel. According to this definition an orbital current density wave is categorized as ``CDW''. In the rest of this subsection we shall first focus on the SDW scattering only.

\begin{figure}[tbp]
\begin{center}
\includegraphics[angle=0,scale=0.45]
{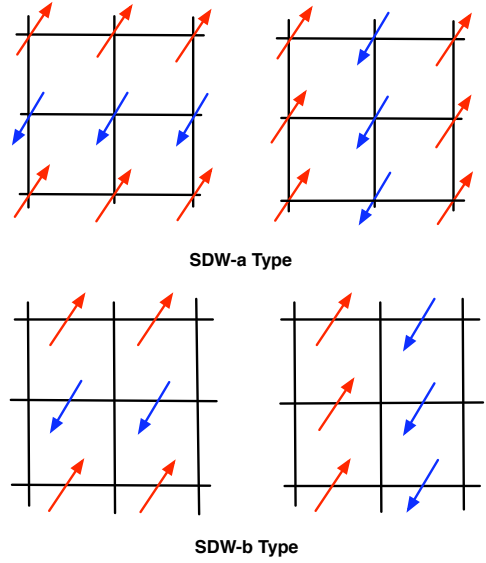}\caption{(color on-line) A schematic real space representation of the two types of SDW order. \label{SDW}}
\end{center}
\end{figure}

The SDW instability already exists at the mean field level in the bare Hamiltonian level. For the type of interaction considered in \Eq{model_5band} the representative pattern of the mean-field magnetic order is shown as the type-a SDW in \Fig{SDW}. It has been shown in Ref.\cite{Ying} that due to the point group symmetry of the band wave function, the off-diagonal SDW matrix element between the larger hole pocket (labeled by 2 in \Fig{FS}) and electron pocket (labeled by 4 and 5) vanishes on the principal axes of the square lattice. As a result the form factor has a $d$-wave symmetry and the SDW state is always gapless. This feature is reproduced by the FRG. More specifically, the leading channel SDW ,i.e., the channel associated with the most negative $w_i$, has a d-wave form the form factor as shown in Fig. \ref{Formfactor}(a). Note here, since the SDW order parameter involves two Fermi surfaces, each form factor in \Fig{Formfactor}(a) is labeled by a pair of FS indices.   For a range of doping, this type of SDW  is the dominant ordering tendency at low energies. In addition, FRG shows a subdominant SDW order. The real space representative is shown as the type-b SDW in \Fig{SDW}. In this pattern the magnet moment is mainly associated with the bonds.

\subsection{The Superconducting Pairing and its Mechanism}

Here we focus on ${\bf k_1}=-{\bf k_2}={\bf k}$ and ${\bf k_3}=-{\bf k_4}={\bf p}$ in reference to \Eq{vf}. These process can be expressed as the sum of singlet and triplet pair scattering
\begin{equation}
V_{\text{SC},\text{s}}({\bf k},{\bf p},\Lambda)\Delta^\dag_{s,{\bf k}}\Delta_{s,{\bf p}}+V_{\text{SC},\text{t}}({\bf k},{\bf p},\Lambda)\Delta^\dag_{t,{\bf k}}\Delta_{t,{\bf p}}.
\end{equation}
Here
\begin{eqnarray}
&&V_{\text{SC},\text{s}}({\bf k},{\bf p},\Lambda)\nn&&=V({\bf k},{\bf -k},{\bf p},{\bf -p},\Lambda)+V({\bf k},{\bf -k},{\bf -p},{\bf p},\Lambda)\nonumber\\ &&+V({\bf -k},{\bf k},{\bf p},{\bf -p},\Lambda)+V({\bf -k},{\bf k},{\bf -p},{\bf p},\Lambda)\label{SCvertex}
\end{eqnarray}
\begin{eqnarray}
&&V_{\text{SC},\text{t}}({\bf k},{\bf p},\Lambda)\nn&&=V({\bf k},{\bf -k},{\bf p},{\bf -p},\Lambda)-V({\bf k},{\bf -k},{\bf -p},{\bf p},\Lambda)\nonumber\\ &&-V({\bf -k},{\bf k},{\bf p},{\bf -p},\Lambda)+V({\bf -k},{\bf k},{\bf -p},{\bf p},\Lambda),
\end{eqnarray}
and
\be
&&\Delta^\dag_{s,{\bf k}}=\frac{1}{\sqrt{2}}\left(\psi^\dag_{s,{\bf k}}\psi^\dag_{s^\prime,{\bf -k}}+\psi^\dag_{s,{\bf -k}}\psi^\dag_{s^\prime,{\bf k}}\right)\nn &&\Delta^\dag_{t,{\bf k}}=\frac{1}{\sqrt{2}}\left(\psi^\dag_{s,{\bf k}}\psi^\dag_{s^\prime,{\bf -k}}-\psi^\dag_{s,{\bf -k}}\psi^\dag_{s^\prime,{\bf k}}\right).
\ee

\begin{figure}[tbp]
\begin{center}
\includegraphics[angle=0,scale=0.45]
{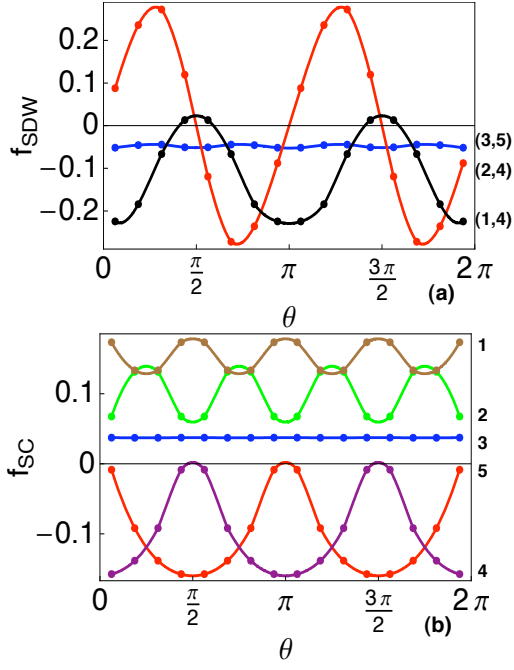}\caption{(color on-line)(a) The FRG form factors of the SDW order parameter. The red , black and blue curves are the form factors associated with the $(2,4)$, $(1,4)$ and $(3,5)$ Fermi pockets (i.e. when $(\v k,\v k+\v Q)\in (2,4),(1,4),(3,5)$ FS), respectively.   (b) The FRG form factors of the SC order parameter. Here different colored curves represent the order parameters associated with the five different FS (i.e. when $(\v k,-\v k)\in 1,...,5$ FS). For each pocket $\theta$ is defined as the angle sustained by the vector connecting the center of the pocket to point $(k_x,k_y)$ on the FS and the vector $(\pi,0)$. The $\v k$-coordinate of the center of FS are $(0,0)$ for pocket 1,2, $(\pi,\pi)$ for pocket 3, $(0,-\pi)$ for pocket 4 and $(\pi,0)$ for pocket 5). The parameters under which the above results are obtained are specified in the text.  \label{Formfactor}}
\end{center}
\end{figure}

For all the parameter we have looked at the triplet pairing tendency is always much weaker than that of the singlet. Among the singlet pairing, we find that the leading mode has $s_{\pm}$ symmetry, i.e. the pairing order parameter does not change sign around each FS, but takes opposite sign between hole band ($\alpha$ and $\beta$) and electron band ($\gamma$). The associated form factor is shown in Fig. \ref{Formfactor}(b)\cite{Fa}. Unlike the SDW, each FS has one form factor.

\begin{figure}[tbp]
\begin{center}
\includegraphics[angle=0,scale=0.48]
{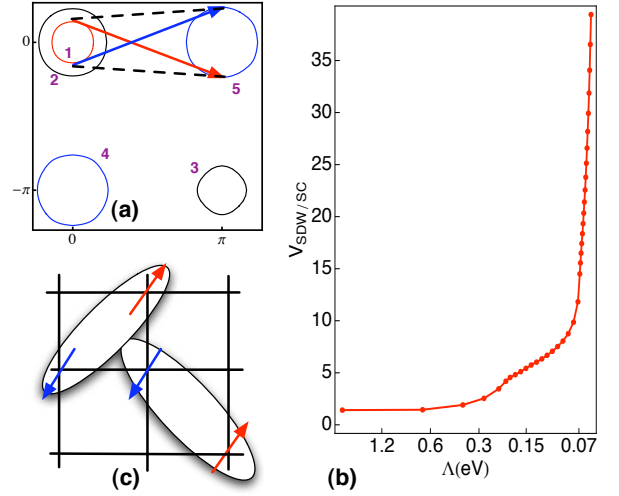}\caption{(color on-line) (a) A schematic representation of the Umklapp scattering processes which has the dual character of SDW and SC. (b) The amplitude of the the scattering process shown in (a) as a function of the FRG energy cutoff $\Lambda$. (c) A schematic real space representation of $\cos k_x\cos k_y$ singlet pairing. \label{SC}}
\end{center}
\end{figure}

One big advantage of FRG is that it allows one to monitor the growth of different classes of scattering processes as a function of the energy cutoff, so it is possible to identify which class grows strong first and hence potentially serves as the driving mechanism for other classes of scattering. As discussed in Ref. \cite{mechanism} we always find the SDW scattering (this includes a whole class of processes) grows strong the first. This is related to the fact that even the bare Hamiltonian shows a SDW mean-field instability. What subsequently triggers the SC scattering is a restricted set of scattering processes which have dual character of SDW and SC scattering as shown in Fig. \ref{SC}(a) \cite{mechanism}. In terms of the scattering Hamiltonian these special type of scattering processes are
\begin{eqnarray}
V({\bf k+Q},{\bf -k+Q},{\bf k},{\bf -k},\Lambda)\psi^\dag_{{\bf k+Q},s}\psi^\dag_{{\bf -k+Q},s^\prime}\psi_{{\bf k},s^\prime}\psi_{{\bf -k},s}.\nonumber\\
\label{shareSC}
\end{eqnarray}
The dual character of this type of scattering can be seen as the possibility of ``factor'' it into both the SDW and SC channels. Together with other SDW scattering, this type of interaction grows positively large in early stages of the FRG (\Fig{SC}(b))\cite{mechanism}. Upon SC factorization this scattering give rise to (positive number)$\times \Delta^\dag_{{\bf k+Q}}\Delta_{{\bf k}}$, hence favors a SC order parameter satisfies
\begin{equation}
\langle\Delta_{{\bf k}}\rangle \langle\Delta^\dag_{{\bf k+Q}}\rangle<0\label{SC_condition}
\end{equation}

Once the above special type of Cooper scattering is seeded, other members of the inter-pocket Cooper scattering processes grow strongly positive in subsequent FRG steps. Two real space representatives of the pairing satisfying \Eq{SC_condition} is are the two types of second neighbor pairing (Fig.\ref{SC}(c)) with form factor $\cos k_x\cos k_y$ and $\sin k_x \sin k_y$. (Of course there are higher angular momentum pairing that satisfies Eq. \ref{SC_condition}, however they in general produce more nodes on the FS and hence are less favorable.) The former has $s_\pm$ symmetry and the latter has $d_{xy}$ symmetry. However, given the locations of the FS of the iron pnictide, the latter is less favorable because it produces four nodes on each FS while the former does not. It is interestingly that precisely these two types of pairing symmetry are observed as the leading ($s_\pm$) and one of less divergent ($d_{xy}$) SC pairing instabilities\cite{Fa}.

The second neighbor singlet pairing in Fig.\ref{SC}(c) is also consistent with the short-range magnetic correlation shown in Fig.\ref{SDW}(a). Indeed according to \Fig{SDW} the next-nearest pairs always preferred to be anti-ferromagnetically correlated. This is in harmony with the singlet pairing depicted in \Fig{SC}(c). From our calculation, the orbital content of the Cooper pair is dominated by $d_{XZ(YZ)}-d_{XZ(YZ)}$, $d_{XZ(YZ)}-d_{YZ(XZ)}$, and $d_{XY}-d_{XY}$.

The SC gap function determined from our FRG analysis often has large variation in the electron pocket. For instance, in Fig. \ref{Formfactor}(b) the gap nearly touches zero along one of the principle axes ($(0,0)-(\pi,0)$) around the electron FS. Such large gap variation is shown to give rise to power law-like NMR relaxation rate so long as the temperature is not smaller than the minimum gap\cite{NMR}. It can also be consistent with the thermal transport experiments results obtained recently\cite{Pong}. The origin of such gap variation is the variation of the orbital content of the band wave function around the electron pocket \cite{anisotropy}. According to band structure\cite{kuroki} the electron band eigenfunction along  $(0,0)-(\pi,0)$ is primarily $XY$-like. In contrast along  $(\pi,0)-(\pi,\pi)$ direction the electron band eigenfunction is dominated by the $XZ$ and $YZ$ orbitals. So if the orbital content of the pairing function is such that $XZ$($YZ$) have more weight than the $XY$ orbital, the type of gap variation will result. Moreover this would also predict that the gap associated with the two hole pockets around $(0,0)$ (labeled by $1$ and $2$) are larger that that of the $(\pi,\pi)$ hole pocket (labeled by $3$) because the former are mainly made of $XZ$ and $YZ$ orbitals and the latter is mostly $XY$-like. This is indeed consistent with the FRG result. The above interpretation of the gap variation is consistent with a recent FRG calculation where the orbital dependence of the interaction matrix element is ignored \cite{Hanke}. Their resulting gap function is far more uniform than ours.

If by tuning some bare interaction parameters, one can reverse the weight of pairing function on different orbitals (i.e. to make the $XY$ orbital weight more than the $XZ$ and $YZ$) it is possible to reverse the gap anisotropy discussed above. In addition, we also suspect that, on the sample surface where the four-fold rotation symmetry is broken, the orbitals content of the band wavefunction will be modified. This could give rise to a different(less) degree of gap anisotropy. If so this could reconcile why surface sensitive measurement such as ARPES observed a far more isotropic gap\cite{DingHong}.
\subsection{The Pomeranchuk Instability}

\begin{figure}[tbp]
\begin{center}
\includegraphics[angle=0,scale=0.45]
{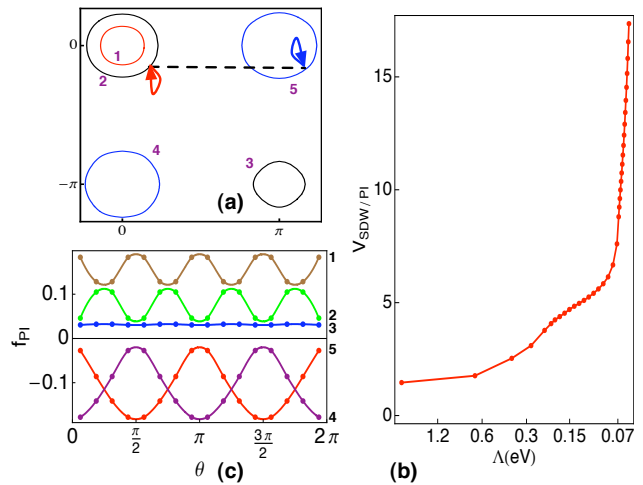}\caption{(color on-line) (a) A schematic representation of the scattering processes which have the dual character of SDW and PI. (b) The amplitude of the the scattering process shown in (a) as a function of the FRG energy cutoff $\Lambda$. (c) The form factor, $f_{{\bf k}}$, in Eq. (\ref{PI_decouple}) associated with the most negative $w_i$. The angle $\theta$ is defined the same way as that in Fig. \ref{Formfactor}   \label{PI}}
\end{center}
\end{figure}

In this section we focus on ${\bf k_1}={\bf k_4}={\bf k}$, ${\bf k_2}={\bf k_3}={\bf p}$ or ${\bf k_1}={\bf k_3}={\bf k}$, ${\bf k_2}={\bf k_4}={\bf p}$ in reference to Eq. \ref{vf}. Similar to Eq. \ref{SDWvertex} and Eq. \ref{cd} we can decompose these forward scattering into the singlet and triplet channels as
\begin{equation}
V_{\text{PI},\text{s}}({\bf k},{\bf p},\Lambda)=V({\bf k},{\bf p},{\bf p},{\bf k},\Lambda)-\frac{1}{2}V({\bf k},{\bf p},{\bf k},{\bf p},\Lambda)
\end{equation}
and
\begin{equation}
V_{\text{PI},\text{t}}({\bf k},{\bf p},\Lambda)=-\frac{1}{2}V({\bf k},{\bf p},{\bf k},{\bf p},\Lambda),
\end{equation}
These channel decoupling allows the forward scattering to be written in the following Fermi liquid form
\begin{equation}
V_{\text{PI},\text{s}}({\bf k},{\bf p},\Lambda)n_{{\bf k}}n_{{\bf p}}+V_{\text{PI},\text{t}}({\bf k},{\bf p},\Lambda)\vec{S}_{{\bf k}}\vec{S}_{{\bf p}}
\end{equation}
where
\begin{equation}
n_{{\bf k}}=\sum\limits_s \psi^\dag_{s{\bf k}}\psi_{s{\bf k}}
\end{equation}
and
\begin{equation}
\vec{S}_{{\bf k}}=\sum\limits_s \psi^\dag_{s,{\bf k}}\vec{\sigma}_{ss^\prime}\psi_{s^\prime,{\bf k}}
\end{equation}

As before we can decompose $V_{\text{PI},{\text{s,t}}}({\bf k},{\bf p},\Lambda)$ into eigen modes as
\begin{eqnarray}
V_{\text{P},{\text{s,t}}}({\bf k},{\bf p},\Lambda)=\sum\limits_{i}w_i(\Lambda) f^i({\bf k})f^i({\bf p}).\label{PI_decouple}\ee

When none of the $w_i$ are negative the FS is stable.  However, when certain $w_i$ become sufficiently negative (so that it can overcome the stiffness imposed by the local Fermi velocity) FS deformation occurs. Attraction in $V_{\text{PI},\text{s}}$ will drive spin-independent FS deformation while attraction in $V_{\text{PI},\text{t}}$ drives spin-depedent FS deformation. In our FRG calculation, we find $V_{\text{PI},\text{t}}$ is usually weaker than $V_{\text{PI},\text{s}}$. As a result, we shall focus on the spin-independent PI in the rest of the subsection.

According to our FRG result, the strongest forward scattering occurs between quasiparticles on the electron and hole pockets. Analogous to the inter-pocket Cooper scattering, there is a special set of such forward scattering processes that have the dual character of being SDW scattering, namely,
\begin{equation}
V({\bf k+Q},{\bf k},{\bf k},{\bf k+Q},\Lambda)\psi^\dag_{{\bf k+Q},s}\psi^\dag_{{\bf k}s^\prime}\psi_{{\bf k}s^\prime}\psi_{{\bf k+Q}s},\label{sharePI}
\end{equation}
as shown in Fig. \ref{PI}(a). Together with other SDW scattering this special type of forward scattering grows strongly positive upon FRG, as shown in Fig. \ref{PI}(b). Viewed as forward scattering this scattering favors the FS deformation at $\v k$ and $\v{k+Q}$ to have opposite sign (due the positive scattering amplitude) . Physically this amounts to the transfer of quasiparticles from ${\bf k}$ to ${\bf k+Q}$ or vice versa. In terms of Eq \ref{PI_decouple} this amounts to
\begin{equation}
f({\bf k})f({\bf k+Q})<0.\label{gg}
\end{equation}
Similar to the ``seeding effect'' discussed in the last subsection, this special set of forward scattering triggers  other forward scattering processes to grow in subsequent renormalization group steps. According to our FRG analysis the $f^i({\bf k})$ associated with the most negative $w_i$ also has $s_{\pm}$ symmetry (see Fig. \ref{PI}(c)), consistent with the requirement of \Eq{gg}.

\begin{figure}[tbp]
\begin{center}
\includegraphics[angle=0,scale=0.45]
{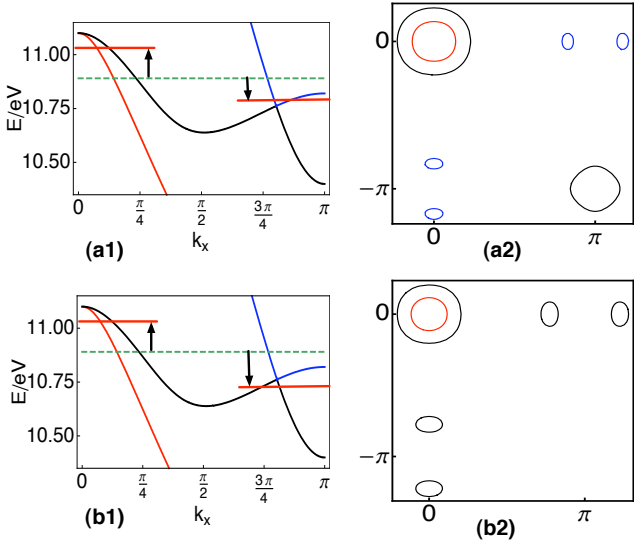}\caption{(color on-line) (a) The FS distortion triggered by the PI. After the distortion each electron pockets is split into two smaller electron pockets. (b) For very strong forward scattering each electron pockets turns into two hole pockets after the FS distortion. In addition the smaller hole pocket centered around $(\pi,\pi)$ is gone. \label{PIeffect}}
\end{center}
\end{figure}

A strong $s_{\pm}$-channel forward scattering favors the transfer of electron  from the electron to the hole pockets or vice versa. As a result, it causes the electron and hole pockets to either both shrink or both expand. The band curvature further determines which possibility is energetically more favorable. As the pockets expand, the Fermi velocity gets larger, hence the kinetic energy cost increases with the deformation amplitude. The reverse is true for the shrinking Fermi pockets. As the result the band curvature always favor both electron and hole pockets to shrink. For a semi-metal with an even number of electrons per unit cell, and simple quadratic dispersing bands, strong repulsive inter-pocket forward scattering will trigger a transition to the band insulator.

The situation for the iron pnictides is more complex due to the band structure. Slightly beneath each of the electron FS there is a Van Hove singularity. A little bit below that there are two Dirac points (two for each electron pocket). In the presence of these band singularities the PI discussed above can induce a topological change in the shape of the electron pockets. As the $w_i$  associated with the $s_\pm$ forward scattering channel gets more negative, a first order transition where the ``Fermi energy" (the {\it band} energy separating filled and empty states after the distortion) in the $\gamma$ band jumps to the vicinity of the Van Hove singularity.
 However, this does not mean a different chemical potential for the electron and hole bands, because the fermi liquid interaction energy also needs to be taken into account. For still more negative $w_{s_\pm}$ all pockets monotonically shrink as a function of decreasing $w_{s_\pm}$.
Before the  $\gamma$-band (the blue curve in Fig. \ref{PIeffect}(a1)) completely empties out,  two electron like pockets will appear near each $M$ point as shown in Fig. \ref{PIeffect}(a2). For the parent compound
 strongly enough inter-pocket forward scattering can empty out all the pockets and hence induce a transition to the band insulator. For the hole doped case,
when the $\gamma$-band becomes completely empty, there are still two sizable hole pockets near the $(0,0)$ point. As the forward scattering gets even stronger, the portion of the $\beta$-band (black curve in Fig. \ref{PIeffect}(b1)) beneath the Dirac points can be depopulated. When that happens two hole-like Fermi pockets will appear near each $(\pi,0)$ or $(0,\pi)$  point as shown in the left of Fig. \ref{PIeffect}(b2). The possibility of an interaction-driven modification of the electron pocket topology has also been studied recently in Ref. \cite{JXLi} where a three-band model and the ``FLEX" approximation were employed.

\begin{figure}[tbp]
\begin{center}
\includegraphics[angle=0,scale=0.45]
{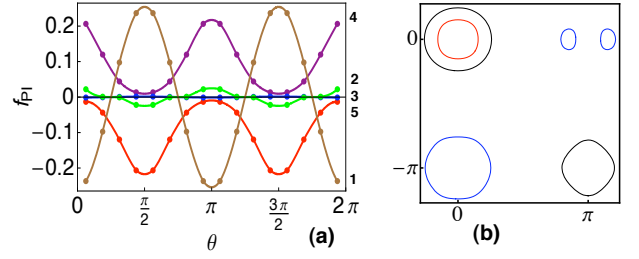}\caption{(color on-line)(a) The form factor, $f_{{\bf k}}$, in Eq. (\ref{PI_decouple}) associated with the sub-leading $w_i$ of Pomeranchuk instability. The angle $\theta$ is defined the same way as that in Fig. \ref{Formfactor};  (b) The FS distortion triggered by the $d_{x^2-y^2}$-type PI. After the distortion one of the electron pockets is better nested with the hole pocket.\label{PI_d}}
\end{center}
\end{figure}

\subsection{A $C_{4v}$ Breaking Pomeranchuk Distortion and its Interaction with the AFM Order}

In addition to the dominant $s_\pm$ distortion channel, our FRG also predicts a sub-leading channel that has $d_{x^2-y^2}$ symmetry, as shown in Fig. \ref{PI_d}(a). {\it This is a band version of the orbital order}. The distorted FS breaks the crystal $C_{4v}$ symmetry. For example, since the deformation of  the two electron pockets have opposite sign, one pocket will expand and the other will shrink. As for the hole pockets, it will expand in one direction and shrink in the orthogonal direction, which makes the hole pocket slightly elongated. When this happens, one out of the two electron pockets will be better nested with the hole pocket, as shown in Fig. \ref{PI_d}(b). As a result a choice among the two possible SDW ordering wavevectors ($(0,\pi)$ and $(\pi,0)$) will be made, and an accompanied lattice distortion is expected. In fact, tetragonal to orthorhombic  structure transition has been observed in both ``1111" and ``122" compounds\cite{neutron},  and theoretical explanation based on $J_1$-$J_2$ model has been proposed \cite{cenke}. Thus there is good reason to believe that the $C_{4v}$ breaking distortion will set in as the SDW order develops. Recent inelastic neutron scattering experiment has revealed a discrepancy between the spin wave dispersion near $(\pi,\pi)$ and the prediction of the $J_1-J_2$ model \cite{pengcheng}. In particular, while the $J_1-J_2$ model predicts a minimum, the actual spin wave dispersion exhibits a maximum at $(\pi,\pi)$. This discrepancy reveals the fact that the $90-$ degree rotation symmetry breaking in the magnetic state has a substantial effect on the electronic structure which in turn modify the magnetic excitation spectrum. In our picture once the SDW-driven $C_{4v}$ breaking FS distortion occurs, the electronic structure become asymmetric with respect to $90-$ degree rotation. This distortion lifts the degeneracy between the $(\pi,0)$ and $(0,\pi)$ orders. So long as the asymmetry is sufficiently big we would expect the minimum in the spin wave dispersion at $(\pi,\pi)$ to turn into a maximum.

\begin{figure}[tbp]
\begin{center}
\includegraphics[angle=0,scale=0.45]
{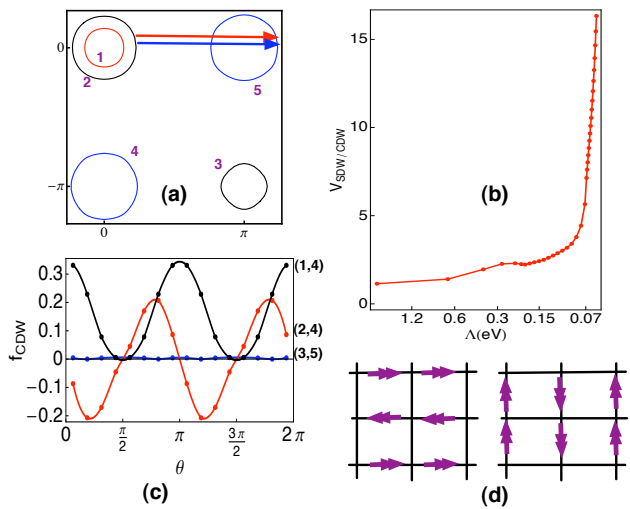}\caption{(color on-line) (a) A schematic representation of the scattering processes which have the dual character of SDW and CDW.
(b) The amplitude of the the scattering process shown in (a) as a function of the FRG energy cutoff $\Lambda$. (c) The form factor $f_{{\bf k}}$ associated with the leading CDW instability. The angle $\theta$ is defined the same way as that in Fig. \ref{Formfactor}; 
(d) A schematic real space representation of orbital current density wave order.  \label{CDW}}
\end{center}
\end{figure}

\subsection{The Orbital Current Density Wave (the Imaginary part of CDW) Order}

As stressed in Sec.II (B) by ``charge density wave'' we mean $(\pi,0)$ or $(0,\pi)$ singlet density wave. This include both the conventional charge density wave with charge density modulation or orbital current density wave where no charge modulation is present.

There is no CDW instability in the bare Hamiltonian. Upon FRG, negative $w_i$s associated with the CDW channels are generated. Interestingly the leading CDW instability is always of the orbital current type. Similar to the Cooper and forward scattering, CDW also share a set of dual scattering processes with the SDW:
\begin{equation}
V({\bf k},{\bf k},{\bf k+Q},{\bf k+Q},\Lambda)\psi^\dag_{{\bf k+Q},s}\psi^\dag_{{\bf k+Q}s^\prime}\psi_{{\bf k}s^\prime}\psi_{{\bf k}s},\label{shareCDW}
\end{equation}
as schematically shown in Fig. \ref{CDW}(a). Like all dual character scattering amplitudes this type of scattering also become strongly positive upon FRG (Fig. \ref{CDW}(b)). This positive scattering favors
\be
d_{\bf k}d^\dagger_{\bf k+Q}<0,\label{cdwc}\ee where \be d_{{\bf k}}=\sum_{s} \psi^\dag_{{\bf k+Q}s}\psi_{{\bf k}s}.\ee Replace $\v k$ by $\v k+\v Q$ in \Eq{cdwc} it is simple to show that $d^\dag_{{\bf k+Q}}=d_{{\bf k}}$.
Combining this result with \Eq{cdwc} we conclude that \be d_{{\bf k}}d_{{\bf k}}<0,\ee as the result an imaginary CDW order parameter $\langle d_{{\bf k}}\rangle$ is favorable. This implies the CDW is of the orbital current type.  The form factor of the most divergent CDW channel is shown in Fig. \ref{CDW}(c) and
a real space representative of this type of CDW order is illustrated in Fig. \ref{CDW}(d). Similar to the SDW case, the relation between the momentum space form factor and the real space pattern is not transparent. This is due to the complex momentum dependance of the Bloch wavefunctions.  The possibility of CDW order has also been recently noted in Ref [\cite{chubukov2}].

\subsection{Competing Order - the Grand FRG Flow }

\begin{figure}[tbp]
\begin{center}
\includegraphics[angle=0,scale=0.42]
{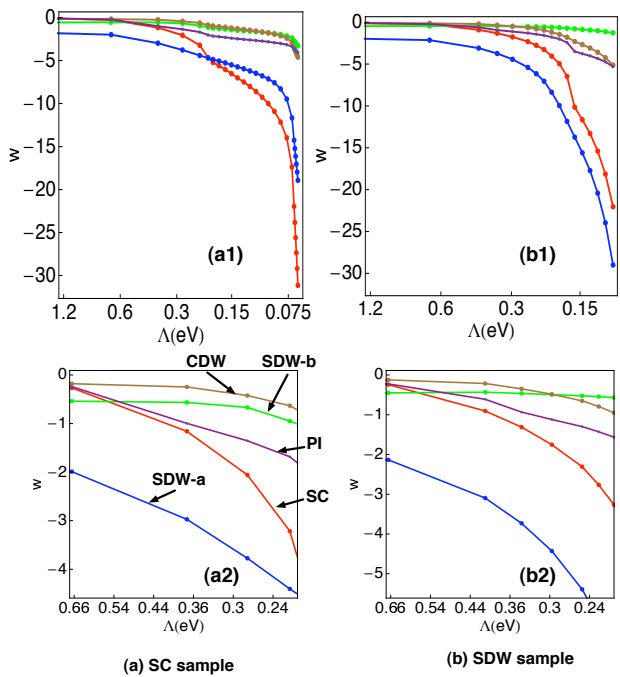}\caption{(color on-line) The grand FRG flow for the iron pnictides. Here the leading $w_i(\Lambda)$ for the SDW,SC,PI and CDW channels are plotted as a function of $\Lambda$. (a) is constructed using the parameters describing a SC sample and (b) is the corresponding plot for a SDW sample. (a1),(b1) cover the full energy scale and (a2),(b2) zoom-in at the intermediate energy regime, with the competing order manifested. \label{Flow}}
\end{center}
\end{figure}

In Fig. \ref{Flow} we plot the $w_i(\Lambda)$ (in reference to \Eq{ord} and \Eq{dec}) for all the leading scattering channels discussed in the last four subsections. We separate the flow into the high, intermediate and low energy regimes.  In the high energy regime (i.e. $\Lambda\sim$ bandwidth, e.g., $\Lambda>1.0eV$), the effective Hamiltonian is very close to the bare Hamiltonian in \Eq{model_5band}. Here the only instability is in the a-type SDW channel (the blue line in Fig. \ref{Flow}(a) and(b)). In the intermediate energy range, (e.g., $0.15eV<\Lambda<1.0eV$), a number of different ordering tendencies are in competition with one another. As shown in Fig. \ref{Flow}(a2) and (b2), this includes different types of SDW, SC, PI and CDW discussed in the last four subsections. In this energy range their relative $w_i$ are dependent on the initial conditions, such as the bare interaction and doping. However, the fact that they are all present at the intermediate energy is very robust. In the lowest energy range, (e.g. $\Lambda<0.15eV$), the a-type SDW (blue line) and the $s_\pm$ SC pairing (red line) dominate over all others. Depends on the doping we show two cases where the SC pairing is the most ``divergent'' scattering in Fig. \ref{Flow}(a) and the SDW scattering is the most divergent in \Fig{Flow}(b). The form factors for these two dominant order tendencies have been shown in Fig. \ref{Formfactor}.

Thus like the cuprates (see Sec.III) iron pnictides exhibit competing order at intermediate energy scales. More specifically both system exhibit the same set of ordering tendencies, i.e., AFM, SC, PI, and CDW. As discussed in the introduction we view this group of competing order as the hallmark of a class of strongly correlated system governed by short-range repulsive interactions.

\subsection{The validity of the $J_1-J_2$ Exchange as the Effective Interaction for the Pnictides}

In Ref.\cite{JP}, much credit to the authors, it is proposed that the $J_1-J_2$ magnetic exchange interaction
\begin{equation}
V_{J_1-J_2}=J_1\sum\limits_{\langle ij\rangle}\vec{S}_i\vec{S}_j+J_2\sum\limits_{\langle\langle ij\rangle\rangle}\vec{S}_i\vec{S}_j,
\end{equation}
captures the AFM and SC correlation of the pnictides.  In the above  $\langle ij\rangle$ represents the nearest neighbor bonds, and $\langle\langle ij\rangle\rangle$ represents the next nearest neighbor bonds. The spin operator expressed in terms of the lattice electron operator is given by  $\vec{S}_i=\sum_{a,s,s^\prime}c^\dag_{ias}\vec{\sigma}_{ss^\prime}c_{ias^\prime}$ where $\vec{\sigma}$ are the three Pauli matrices and $a$ is the  orbital index.
After Fourier transform this gives rise to the following electron-electron scattering Hamiltonian in the momentum space
\begin{equation}
J_2 V_{J_1-J_2}\left({\bf k_1},{\bf k_2},{\bf k_3},{\bf k_4},\frac{J_1}{J_2}\right)\psi^\dag_{{\bf k_1},s}\psi^\dag_{{\bf k_2},s^\prime}\psi_{{\bf k_3},s^\prime}\psi_{{\bf k_4},s},\label{j12}
\end{equation}
where we have factored out an overall energy scale $\sim J_2$.

In this subsection we test whether such an interaction reproduces the dominant low energy interaction generated by our FRG.  The minimum requirement is for \Eq{j12} to capture the lowest energy SDW and Cooper scattering. More specifically, we regard \Eq{j12} as the fully renormalized interaction in \Eq{vf} and extract the scattering in various channels discussed in subsection IIA-IID. We then decompose the associated scattering into eigen channels as in Eq. \ref{dec}. We check whether (i) the $w_i$ associated with the SDW and SC scattering are more negative than all other channels, and (ii) whether the symmetry of the leading SDW and SC form factor agrees with that of the a-type SDW and $s_\pm$. If both (i) and (ii) are passed we compute the overlap between the form factor predicted by the $J_1-J_2$ model and our FRG result. In practice, we use the same set of discrete $\v k$ points on FS to determine the form factor of the $J_1-J_2$ interaction as used in our FRG calculation.

For $-1.5J_2<\sim J_1 <\sim J_2$, the $J_1-J_2$ model passes (i) and (ii) and the results for the overlaps are shown in Fig. \ref{fit}. For both SDW and SC the best overlap is reached when a small and negative $J_1$ is allowed. Thus at the lowest energy, the $J_1-J_2$ model passes with flying colors. For the intermediate energy range the  $J_1-J_2$ model can also reproduce the form factor for the spin-independent Pomeranchuk distortion described in subsection II(C). However it can not reproduce the CDW form factor and other subleading SDW,SC and PI form factors.  Until writing we have not succeeded in finding a simple local model that can capture the form factor associated with all channels.

Thus we conclude the $J_1-J_2$ model is sufficient to describe the low energy physics of the iron pnictides where only SDW and SC are important. We do not think it should be used as a lattice model with the full kinetic bandwidth.

\begin{figure}[tbp]
\begin{center}
\includegraphics[angle=0,scale=0.45]
{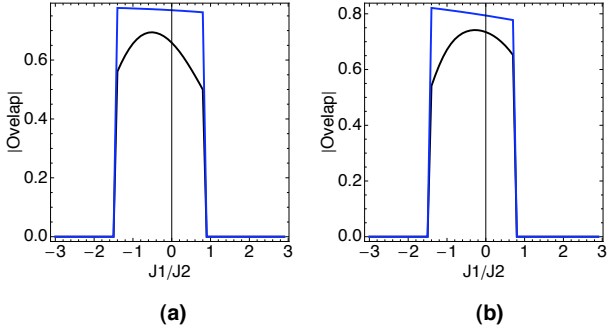}\caption{(color on-line) The overlap between the form factors predicted by the $J_1-J_2$ model and those obtained from the FRG. The black and blue curves are the overlaps associated with the SC SDW form factors, respectively.  (a) and (b) corresponds to SC and SDW samples in Fig. \ref{Flow}. The overlap is set to zero when the symmetry of the form factors do not agree, or when SDW and SC are not the leading instability of the $J_1-J_2$ model . \label{fit}}
\end{center}
\end{figure}

\section{The FRG Results for the One-band $t-t^\prime$ Hubbard Model}

\begin{figure}[tbp]
\begin{center}
\includegraphics[angle=0,scale=0.48]
{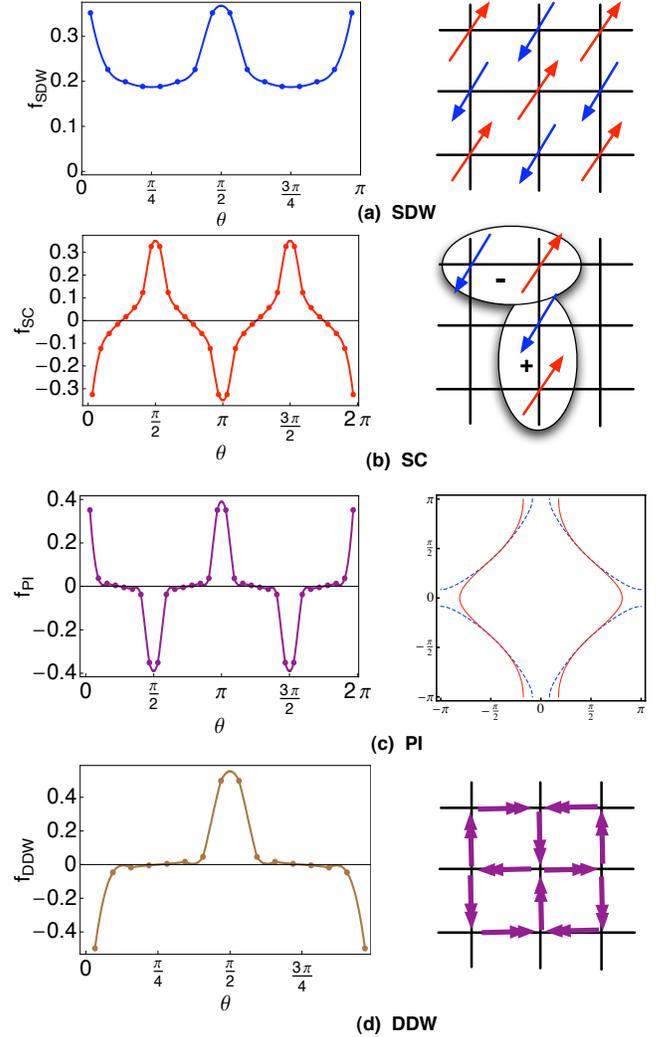}\caption{(color on-line) The FRG results for the $t-t^\prime$ Hubbard model. (a) The SDW form factor and a schematic real space representation of the $(\pi,\pi)$ AFM order. (b) The  $d$-wave SC pairing form factor and its schematic real space represnetation. (c) The PI Form factor and the distorted FS surface. The blue dashed line and the red solid line denote the original and the distorted FS, respectively.  (d) The imaginary CDW, or the orbital current order, form factor, and its schematic real space representation. \label{oneband_orders}}
\end{center}
\end{figure}

In this section we present the FRG results for $t-t^\prime$ single-band Hubbard model. This model was used in Ref.\cite{Honerkamp} to describe the cuprates. It is well known that for the cuprates the Hubbard interaction is a bit too strong for the FRG approach. Here we take a different point of view. Let us imagine a material with exactly the same band structure as the cuprates but with a weaker correlation. Will this material exhibit similar ordering tendencies as the cuprates? If so,  it can be argued that this material is adiabatically connected to the cuprates. Intuitively one expects this type of intermediate coupling system to resemble more the overdoped than the underdoped cuprates. The main purpose of this section is to compare the FRG results for this type of hypothetical system with those of the iron pnictides discussed earlier. The hope is that some hints on the pairing mechanism for both will emerge from such a comparison.

The Hamiltonian we studied here is defined as
\begin{equation}
\mathcal{H}=\sum_{\v k,s}\epsilon_{\v k}n_{\v k s}+\sum\limits_{i}\left(U n_{i\uparrow}n_{i\downarrow}-\mu n_i\right)\label{Hubbard}
\end{equation}
where the band dispersion is
\begin{equation}
\epsilon_{{\bf k}}=-2t(\cos k_x+\cos k_y)+4 t^\prime \cos k_x \cos k_y-\mu.\label{ek}
\end{equation}
Hereafter in unit of $t$ (i.e. $t\equiv 1$) we shall take $t^\prime=0.2$ and $U=4$, we shall set  $\mu=-0.7$ which corresponds to $10\%$ hole doping. The band dispersion $\epsilon_{\v k}$ of \Eq{ek} gives a FS shown as the blue curves in left panels (a)-(c) of \Fig{shareprocess}.

\begin{figure}[tbp]
\begin{center}
\includegraphics[angle=0,scale=0.43]
{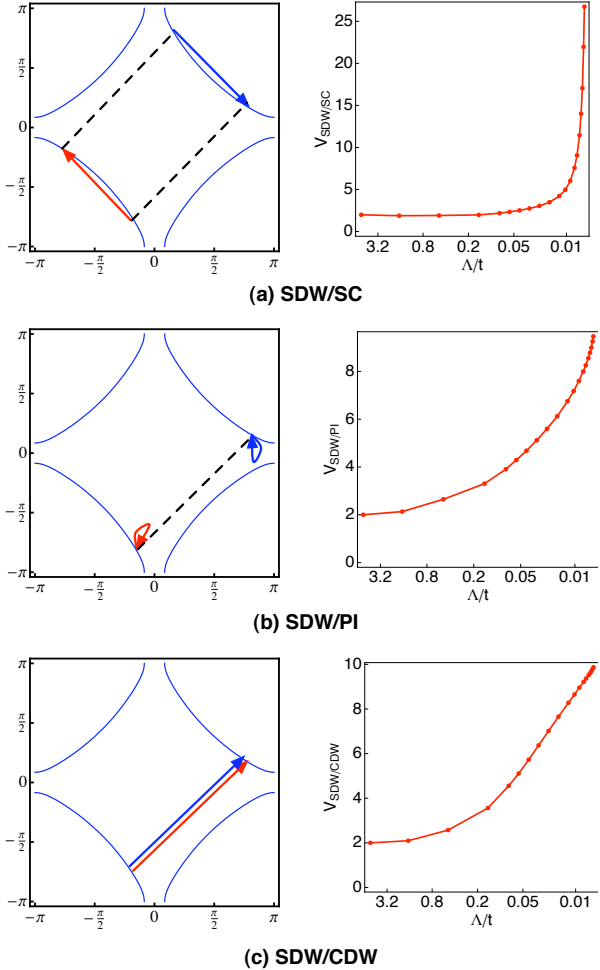}\caption{(color on-line) The dual scattering processes for the one-band $t-t^\prime$ Hubbard model. (a) The dual SDW-SC processes. (b) The dual SDW-PI processes, and (c) The dual SDW-CDW processes. The left column is the schematic of the scattering processes and the right column shows the amplitude of the the scattering process as a function of the FRG energy cutoff $\Lambda$.
\label{shareprocess}}
\end{center}
\end{figure}

\subsection{D-symmetry SC Pairing, FS Distortion and Orbital Current Order}

Upon FRG the $(\pi,\pi)$ SDW and the Cooper scattering grow strong at low energies. Their corresponding leading channel form factors (in reference to \Eq{dec}) are shown in Fig. \ref{oneband_orders}(a,b). The SC pairing factor factor has the well-known $d_{x^2-y^2}$ symmetry.  In addition, the FRG also generates the PI and (imaginary) CDW ordering tendency at low energies. Their associated form factor also have the $d_{x^2-y^2}$ symmetry. The $d_{x^2-y^2}$ PI will trigger a $C_{4v}$-breaking FS distortion as shown in Fig. \ref{oneband_orders}(c) \cite{d-wave}. The above CDW order is the famous ``DDW'' orbital current state (Fig. \ref{oneband_orders}.(d)).

\begin{figure}[tbp]
\begin{center}
\includegraphics[angle=0,scale=0.43]
{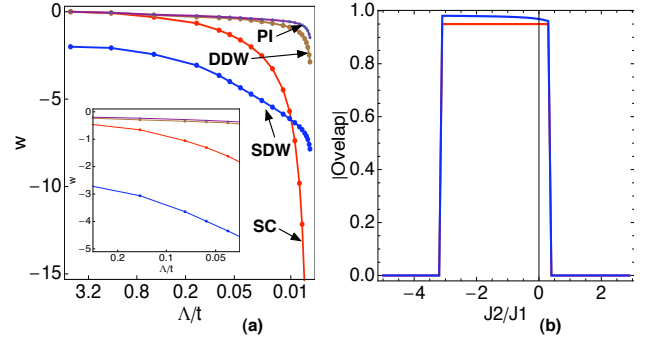}\caption{(color on-line)(a) The grand FRG flow for the $t-t^\prime$ Hubbard model. Here the $w_i(\Lambda)$ associated with the SDW,SC,PI and CDW are plotted. The inset zooms in at the intermediate energy regime. (b) The overlap of the FRG form factors with those predicted by the $J_1-J_2$ model. The red and the blue curves are the overlaps for the SC and SDW form factors, respectively. The overlap is set to zero when the symmetry of form factors do not agree or when the $J_1-J_2$ model does not predict SC and SDW as the leading instabilities.   \label{oneband_Flow}}
\end{center}
\end{figure}

Like the pnictides the strong SC, PI and CDW scattering all grow out of the dual scattering processes shared by them and the SDW scattering . For example, the scattering Hamiltonian describing the dual SC  and SDW scattering is given by Eq. (\ref{shareSC}) (here $\v Q=(\pi,\pi)$) and is illustrated in Fig. \ref{shareprocess}(a). Since such Cooper scattering is shared with SDW, it has a positive amplitude, and favors  $\langle \Delta_{{\bf k}}\rangle\langle \Delta_{{\bf k+Q}}\rangle<0$. Two representative real space pairing form factor whose Fourier transform has this symmetry are given by  $\cos k_x+\cos k_y$ and  $\cos k_x-\cos k_y$.  For small $t^\prime/t$ the node lines of  $\cos k_x+\cos k_y$ ($\cos k_x+\cos k_y=0$) lies very close to the FS and intersects it many times. On the other hand the node lines of $\cos k_x-\cos k_y$ basically runs perpendicular to the FS and only intersects it at four points. As a result the latter pairing symmetry is more energetically favorable.  Similarly, the PI and SDW share the processes of Eq. \ref{sharePI} as shown in Fig. \ref{shareprocess}(b). These repulsive processes favor the transfer quasi-particle from ${\bf k}$ to ${\bf k+Q}$ and yield a FS deformation as shown in Fig. \ref{oneband_orders}(c). CDW and SDW share the process of Eq. (\ref{shareCDW}) and it is shown in Fig. \ref{shareprocess}(c), which gives rise to charge current order for the same reason as discussed in section II (D).

We show in Fig. \ref{oneband_Flow} the leading $w_i$ for the SDW, SC, PI and CDW scattering as a function of the energy cutoff $\Lambda$. Similar to the iron pnictides case only SDW channel has negative $w_i$ for large $\Lambda$. For intermediate $\Lambda$, as shown in the inset of Fig. \ref{oneband_Flow}, the $w_i$ associated with all the scattering channels mentioned above are negative, signaling competing order at intermediate energies. For small $\Lambda$  SC (and SDW) ordering tendencies clearly surpass all others. Following the same strategy as used in Sec.II, we fit the FRG form factor of SC and SDW to those predicted by a $J_1-J_2$ model. As shown in Fig. \ref{oneband_Flow}(b), once $J_2$ becomes positive, the fit worsens immediately. While the fit remains very good for a large range of negative $J_2$ until some other instability becomes dominating. This is because a negative $J_2$ does not introduce any frustration to the $(\pi,\pi)$ SDW correlation.

\subsection{The Stripe Tendency}

The propensity to stripe order has been observed by several experimental probes (e.g., inelastic neutron scattering, and STM) for several families of underdoped cuprates. In particular inelastic neutron scattering has found incommensurate peaks at low energies for LSCO and YBCO \cite{neutron_cup}. At high bias (comparable to the pseudogap energy scale) STM has revealed glassy pattern of local 90-degree rotation symmetry breaking, and a spatial modulating local density of states with a period of $\sim 4-5$ lattice spacing in Bi2212 and NaCCOC \cite{STM_stripe}. Recent ARPES experiments performed on Bi2212 have found the evidences for two different gaps. The gap in the anti-nodal region appears to be particle-hole asymmetric consistent with a density wave gap\cite{zx}. These experimental evidences point to the possibility that the gap in the antinodal region is due to the formation of a density wave, while that in the nodal region is due to SC pairing.

\begin{figure}[tbp]
\begin{center}
\includegraphics[angle=0,scale=0.43]
{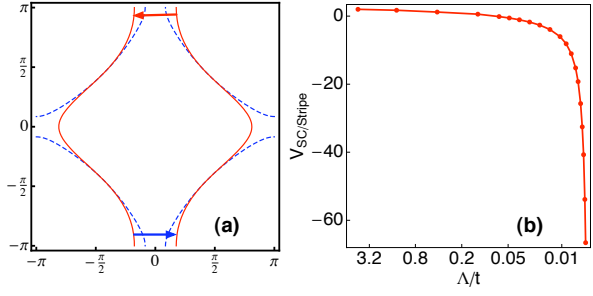}\caption{(color on-line)(a) A schematic illustration of the distorted FS and the scattering processes that have the dual character of SC and incommensurate uni-direction charge density wave, or in short stripe order. The blue solid line is original FS and the red dashed line is the distorted FS triggered by the PI. (b) The FRG flow of the scattering amplitude associated with (a). \label{stripe}}
\end{center}
\end{figure}

In this subsection we examine the possibility of stripe order from the FRG point of view. It has been pointed out early on \cite{XJZhou} that the nearly parallel (hence well nested) FS segments in the antinodal regime could cause the propensity for CDW order. The nesting wavevectors are  $(q,0)$ or $(0,q)$ with $q\sim 2\pi/5$. Interestingly, after the PI distortion, the antinodal FS become better nested (\Fig{stripe}) in one direction and the nesting vector is slightly increased. Interestingly the scattering process shown in Fig. \ref{stripe}(a) is a also Cooper process.  When the $d$-wave Cooper scattering becomes very strong, the amplitudes associated with \Fig{stripe}(a) becomes strongly negative (because it connects pair states with the same gap sign). Upon further FRG the scattering amplitudes associated with these SC/stripe dual processes continuously to grow in magnitudes (Fig. \ref{stripe}(b)). When this type of scattering is decoupled in the CDW channel, due to its negative amplitude, favors a real CDW order parameter. This implies real charge density modulation rather than non-zero orbital current. The uni-directional charge density wave resulted from this could be consistent  with charge stripes.

\section{The Double-Layer Hubbard Model}

As we emphasized in the previous sections, from the FRG point of view, the key prediction of the magnetic scattering driven SC is that the SC order parameter has to satisfy $\Delta_{{\bf k}}\Delta_{{\bf k+Q}}<0$. In general there could be many different pairing functions satisfying this constraint. Finally which pairing function is more energetically favorable is determined by which form factor can give rise to maximal gapping of the FS. In this section, we will further strengthen this point by studying a double-layer Hubbard model which is schematically shown in Fig. \ref{model}. It consists of Hubbard layers coupled by a vertical tunneling $t_z$ \cite{bilayer}. The Hamiltonian is given by
\begin{equation}
\mathcal{H}=\mathcal{H}_1+\mathcal{H}_2-t_z\sum\limits_{i,s}c^\dag_{1,i,s}c_{2,i,s}+h.c.
\end{equation}
Within each layer, it is described by the Hamiltonian $H_{1,2}$ as in Eq. (\ref{Hubbard})(except we set $t^\prime=0$ for this study).

\begin{figure}[tbp]
\begin{center}
\includegraphics[angle=0,scale=0.43]
{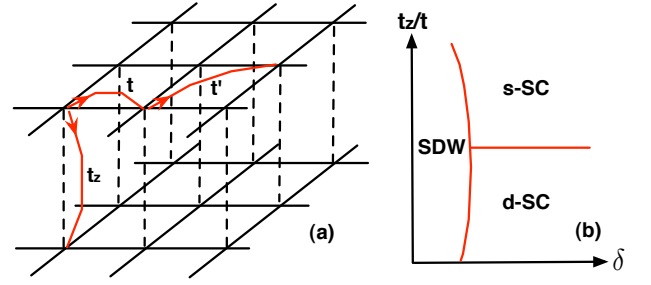}\caption{(color on-line)(a) A schematic illustration of hopping associated with the double payer Hubbard model. (b) A schematic representation of the proposed phase diagram, where $\delta$ is doping. \label{model}}
\end{center}
\end{figure}

\begin{figure}[tbp]
\begin{center}
\includegraphics[angle=0,scale=0.43]
{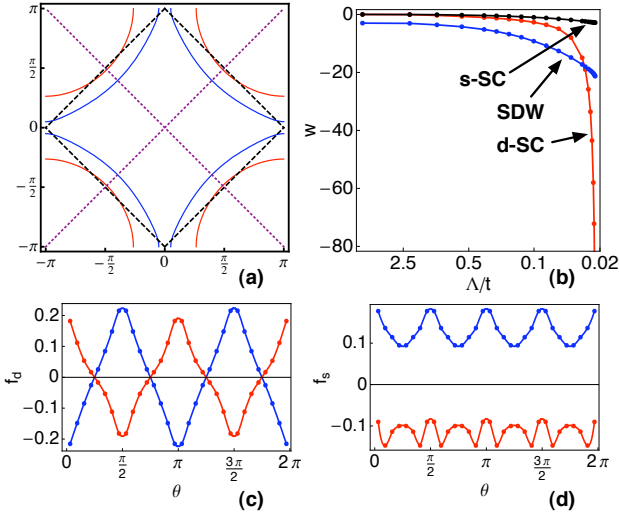}\caption{(color on-line)(a) The FS of the double-layer Hubbard model with $t_z=0.5$ at $4\%$ hole doping.
The two solid lines (blue and red) are the electron and hole FS respectively. The purple dotted line is the nodal line of $\cos k_x-\cos k_y$, and the black dashed line is the nodal line of  $\cos k_x+\cos k_y$.
(b) The $w_i(\Lambda)$ associated with SDW, extended $s$-wave SC and $d$-wave SC. (c) The FRG predicted d-wave pairing form factor. (d) The FRG predicted form factor of the subleading extended $s$-wave pairing. Here for electron FS around $(0,0)$ , $\theta$ is the angle between $(k_x,k_y)$ and $(\pi,0)$. For the hole FS around $(\pi,\pi)$, $\theta$ is the angle between $(k_x-\pi,k_y-\pi)$ and $(\pi,0)$. \label{tz05}}
\end{center}
\end{figure}

\begin{figure}[tbp]
\begin{center}
\includegraphics[angle=0,scale=0.43]
{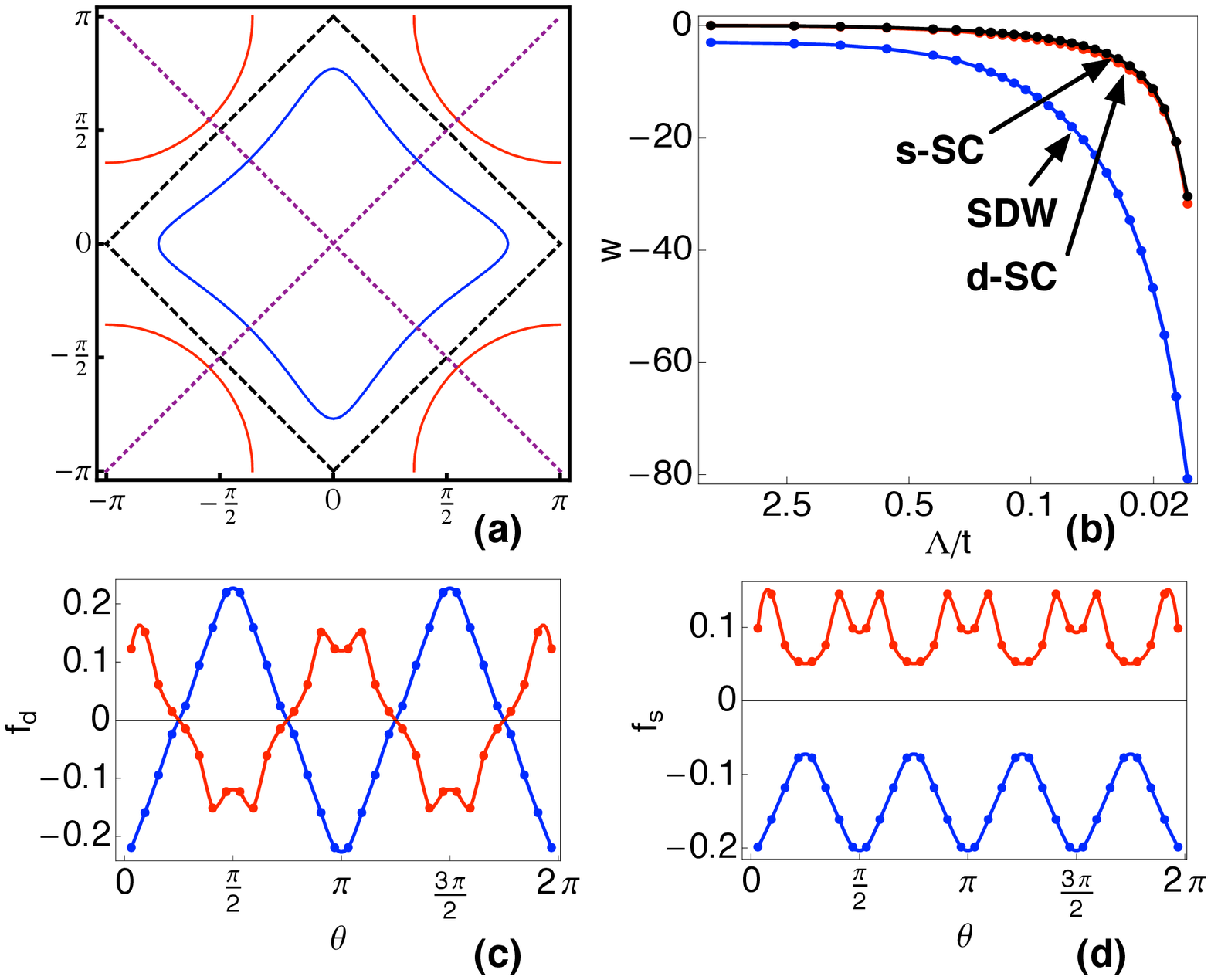}\caption{(color on-line)(a) The FS of the double-layer Hubbard model with $t_z=1.0$ at $3\%$ hole doped
(b) The $w_i(\Lambda)$ associated with SDW, extended $s$-wave SC and $d$-wave SC.
 (c) The FRG predicted d-wave pairing form factor. (d) The FRG predicted form factor of the nearly degenerate extended $s$-wave pairing.
The angle is defined in the same way as \Fig{tz05}. \label{tz10}}
\end{center}
\end{figure}

\begin{figure}[htbp]
\begin{center}
\includegraphics[angle=0,scale=0.43]
{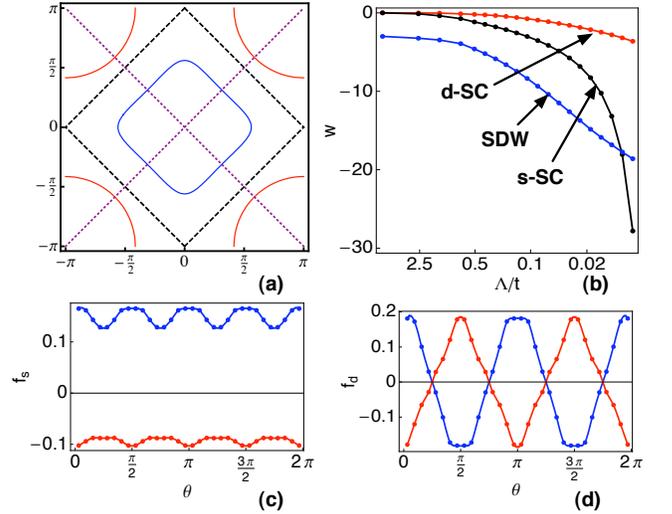}\caption{(color on-line)(a) The FS of the double-layer Hubbard model with $t_z=1.5$ at $8\%$ hole doping.
(b) The $w_i(\Lambda)$ associated with SDW, extended $s$-wave SC and $d$-wave SC.
(c) The FRG predicted extended s-wave pairing form factor. and (d)  (d) The FRG predicted form factor of the less divergent $d$-wave pairing. The angle is defined in the same way as \Fig{tz05}. \label{tz15}}
\end{center}
\end{figure}

With the interlayer hopping, the vertical pair of sites in each unit cell form bonding and anti-bonding combinations as follows
\begin{eqnarray}
d_{1,i,s}=\frac{1}{\sqrt{2}}\left(c_{1,i,s}+c_{2,i,s}\right)\nonumber\\
d_{2,i,s}=\frac{1}{\sqrt{2}}\left(c_{1,i,s}-c_{2,i,s}\right).
\end{eqnarray}
In terms of the $d$ operators the Hubbard interaction becomes
\begin{eqnarray}
&&\frac{U}{2}\sum\limits_{i,s,s^\prime}d^\dag_{1,i,s}d^\dag_{1,i,s^\prime}d_{1,i,s^\prime}d_{1,i,s}+d^\dag_{2,i,s}d^\dag_{2,i,s^\prime}d_{2,i,s^\prime}d_{2,i,s}\nonumber\\
&&+U\sum\limits_{i}d^\dag_{1,i,s}d^\dag_{2,i,s^\prime}d_{2,i,s^\prime}d_{1,i,s}\nonumber\\
&&+U\sum\limits_{i}d^\dag_{1,i,s}d^\dag_{2,i,s^\prime}d_{1,i,s^\prime}d_{2,i,s}\nonumber\\
&&+\frac{U}{2}\sum\limits_{i,s,s^\prime}d^\dag_{1,i,s}d^\dag_{1,i,s^\prime}d_{2,i,s^\prime}d_{2,i,s}+d^\dag_{2,i,s}d^\dag_{2,i,s^\prime}d_{1,i,s^\prime}d_{1,i,s}.\nonumber\\
\end{eqnarray}
The energy bands formed by the bonding and anti-bonding orbitals are given by  $\epsilon({\bf k})\pm t_z$ where
$\epsilon(\v k)=-2t (\cos k_x+\cos k_y)+4 t^\prime \cos k_x\cos k_y-\mu$.  Thus, as long as $t_z$ is smaller than the band width of $\epsilon({\bf k})$, there will be an electron FS around $(0,0)$ and a hole FS around $(\pi,\pi)$. These FS are well nested at half-filling (two electrons per unit cell) since their size are the same. Thus with any non-zero positive $U$ we expect $\v Q=(\pi,\pi)$ SDW order. Away from half filling nesting no longer exists, however we still expect strong $(\pi,\pi)$ antiferromagnetic fluctuations. Using the arguments presented in Sec.II and Sec.III we expect SDW driven superconducting pairing satisfying $\Delta_{\v k}\Delta_{\v k+\v Q}<0$. Both extended s-wave $\cos k_x+\cos k_y$ and  $d_{x^2-y^2}$-wave $\cos k_x-\cos k_y$ form factors satisfy this requirement. In the limit of small $t_z$ the nodes of $\cos k_x+\cos k_y$ lies close to both FS (\Fig{tz05}(a)). As the result $d_{x^2-y^2}$ will be favored. As $t_z$ increases, both the electron and hole pockets shrink. As they moves away from the node line of $\cos k_x+\cos k_y$ pairing (\Fig{tz15}(a)) the extended s-wave pairing will become favored over the $d_{x^2-y^2}$ pairing because the latter still produce nodes on the FS \cite{kuroki_flex}. With the extended s-wave pairing the gap function on the two FS have opposite sign which is reminiscent of the gaps proposed in iron pnictides. We will follow the term used in the iron pnictide and call it $s_{\pm}$ SC. The above argument suggests that as a function of increasing $t_z$ there is a phase transition from $d$-wave SC pairing to the $s_{\pm}$ pairing. A schematic phase diagram is shown in Fig. \ref{model}(b).

In Fig. \ref{tz05},\ref{tz10},\ref{tz15} we present the FRG result for this model for $t_z=0.5,1.0$ and $1.5$. For $t_z=0.5$, as shown in Fig. \ref{tz05}(a), the FS intersects with the nodal line of  $\cos k_x+\cos k_y$ many time. In this case, as shown in Fig. \ref{tz05}(b), the $w_i$ associated with $d$-wave pairing grows much faster than that of $s_{\pm}$-wave. The FRG form factor associated with these two types of pairing are shown in Fig. \ref{tz05}(c) and (d), respectively. For $t_z=1.0$, the FS is shown in Fig. \ref{tz10}(a). Although the nodal line of $\cos k_x+\cos k_y$  no longer intersects the FS, a large portion of the hole FS lies close to the nodal line, hence will have a small gap. On the other hand, the nodal line of $\cos k_x-\cos k_y$ intersect with both two FS four times, giving rise to four nodes on each. It will be hard to tell which one will be favored by the kinetic energy. In fact, as shown in Fig. \ref{tz10}(b), the $w_i$ of both types of pairing are nearly degenerate. As $t_z$ further increases, e.g., $t_z=1.5$ as shown in Fig. \ref{tz15}, the $w_i$ associated with $s_{\pm}$ pairing dominates over all other pairing form factors. In short, our FRG calculation confirms the earlier heuristic argument suggesting a transition between the $d_{x^2-y^2}$ pairing and $s_{\pm}$ pairing. It would be very interesting to fabricate a double material with tunable interlayer hopping strength. For example it might be possible to realize this type of model using fermion optical lattices when the technical problem of cooling is overcome.

\section{The Renormalization Group Study of a Ladder Version of the Iron Pnictides}

In the cuprate physics, the study of the  ladder version of the Hubbard model has contributed significantly to
our understanding. Due to the quasi-one-dimensional nature of the system the Hubbard ladder is amenable to more exact treatment. Examples include density matrix renormalization group (DMRG), perturbative RG plus bosonization etc. Motivated by the above we perform the perturbative RG study for the two-leg ladder version of Eq. \ref{model_5band}. The formulation, terminology and notations used in this section closely follows those of Ref.[\cite{HHLin}]. The readers can consult it for more details.

The greatest simplification of the ladder RG, as compared to the FRG, is the fact that FS become Fermi points. As a result the functional RG equation become a group of coupling constant recursion relations. These equations can be integrated numerically to yield the coupling constants as a function of the energy cutoff $\Lambda$. As we shall show in the following, the qualitative nature of the ladder RG results agree with the two-dimensional FRG ones. In particular we also get the spin singlet $s_{\pm}$ pairing. In addition, the same pairing mechanism applies.
\begin{figure}[tbp]
\begin{center}
\includegraphics[angle=0,scale=0.43]
{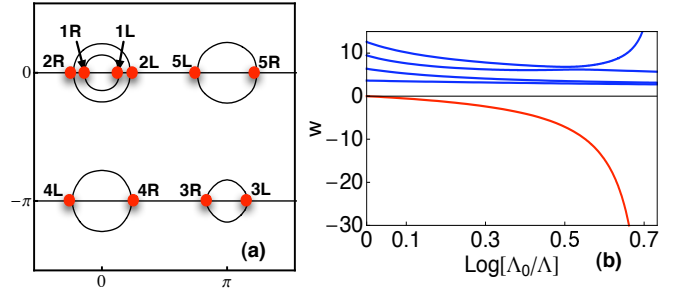}\caption{(color on-line)(a) The FS of iron pnictide and the five pairs of Fermi points for the iron pnictide ladder. (b) The $w_i(\Lambda)$ associated with all singlet SC pairing.  The red line is the $w_i$ associated with the leading SC instability. \label{ladderRG}}
\end{center}
\end{figure}

 In the following we consider a 2-leg ladder running in the $x$ direction. The $y$ direction is consisted of two chains, and periodic boundary condition is applied in both directions. The Hamiltonian we use is the same as Eq. \ref{model_5band}. The only difference is the first Brillouin zone is changed from $-\pi\le k_{x,y}\le \pi$ to $-\pi\le k_x\le \pi,~k_y=0,\pi$. Due to the discrete value of $k_y$ the five FS becomes five pair of Fermi points (\Fig{ladderRG}). Fortunately most of the high symmetry points on the 2D FS are captured. Following Ref.[\cite{HHLin}]
we label these Fermi points as $nR,nL$ where $n=1,\dots,5$ and $R,L$ denote right and left movers respectively.
It is worth to note that for the electron FS $R,L$ correspond to the right and left members of the Fermi pair, and for the hole FS, $R,L$ denotes the left and right members of the Fermi pair. The reversal is due to the fact that the dispersion around the hole FS is the opposite of that around the electron ones. More specifically we have
$1R,1L=(\mp k_{F1},0)$, $2R,2L=(\mp k_{F2},0)$, $3R,3L=(\mp k_{F3},-\pi)$,  $4R,4L=(\pm k_{F4},-\pi)$, $5R,5L=(\pi\pm k_{F5},0)$, where $1,2,3$ label the hole FS while $4,5$ label the electron FS. Associated with each Fermi pair there are the ``slowly varying" chiral fields $\psi_{Rn\alpha}$ and $\psi_{Ln\alpha}$ where $\alpha$ is spin index.
The kinetic part of the Hamiltonian, linearized around the Fermi points, is given by
\begin{equation*}
H_0=\sum_{\alpha}\sum_{n=1}^{5} v_{Fi}
 \int {\rm d}x (
  \psi_{nR\alpha}^\dagger i \partial_{x}\psi_{nR\alpha}^{\vphantom{\dagger}}
- \psi_{nL\alpha}^\dagger i \partial_{x}\psi_{nL\alpha}^{\vphantom{\dagger}}
)
\end{equation*}
where $v_{Fn}$ is the Fermi velocity associated with the $n$-th Fermi pair.

In the following we assume there is {\em no} nesting (hence no Umklapp process exists).
Under this condition, the set of momentum conserving four-point vertices have the same form as those in a
5-leg one band Hubbard ladder at generic dopings\cite{HHLin}.
These interactions are given by
\be
H_{\rm int}=  & \sum_{n,m=1}^{5} (
 \tilde{c}_{nm}^{\rho} J^{R}_{nm} J^{L}_{nm}
 -\tilde{c}_{nm}^{\sigma} \vec{J}^{R}_{nm} \cdot \vec{J}^{L}_{nm}
) \nn
 & +\sum_{n\neq m}(
 \tilde{f}_{nm}^{\rho} J^{R}_{nn} J^{L}_{mm}
 -\tilde{f}_{nm}^{\sigma} \vec{J}^{R}_{nn}\cdot \vec{J}^{L}_{mm}
)
\ee
where
\be
&&J^{R}_{nm}=\sum_{\alpha}\psi_{Rn\alpha}^\dagger
 \psi_{Rm\alpha}^{\vphantom{\dagger}},~\vec{J}^{R}_{nm}=\frac{1}{2}\sum_{\alpha,\beta}\psi_{Rn\alpha}^\dagger
 (\vec{\sigma})_{\alpha\beta} \psi_{Rm\beta}^{\vphantom{\dagger}}\nn&&J^{L}_{nm}=\sum_{\alpha}\psi_{Ln\alpha}^\dagger
 \psi_{Lm\alpha}^{\vphantom{\dagger}},~\vec{J}^{L}_{nm}=\frac{1}{2}\sum_{\alpha,\beta}\psi_{Ln\alpha}^\dagger
 (\vec{\sigma})_{\alpha\beta} \psi_{Lm\beta}^{\vphantom{\dagger}}\nonumber
\ee
with $\vec{\sigma}$ being the three Pauli matrices.

Since symmetry requires $\tilde{c}_{nm}^{\rho,\sigma}=\tilde{c}_{mn}^{\rho,\sigma}$
and $\tilde{f}_{nm}^{\rho,\sigma}=\tilde{f}_{mn}^{\rho,\sigma}$, we have a total of $2\times (15+10)=50$ coupling constants. In addition the ladder RG starts from a very small cutoff, so there is no ``projection error'' as in the FRG\cite{Honerkamp,Fa}.

The RG equations can be easily obtained from Eq. (3.7)-(3.10) of Ref[\cite{HHLin}]
\begin{eqnarray*}
\dot{f}_{nm}^{\rho} & = &
 (c_{nm}^{\rho})^2+\frac{3}{16}(c_{nm}^{\sigma})^2 \\
\dot{f}_{nm}^{\sigma} & = &
 -(f_{nm}^{\sigma})^2+ 2c_{nm}^{\rho}c_{nm}^{\sigma}
 -\frac{1}{2}(c_{nm}^{\sigma})^2 \\
\dot{c}_{nm}^{\rho} & = &
 2 (c_{nm}^{\rho} f_{nm}^{\rho}+\frac{3}{16} c_{nm}^{\sigma} f_{nm}^{\sigma})
\\ & &
 - \sum_{k} \alpha_{nm,k} (c_{nk}^{\rho}c_{km}^{\rho}
 +\frac{3}{16}c_{nk}^{\sigma}c_{km}^{\sigma}) \\
\dot{c}_{nm}^{\sigma} & = &
 (2 c_{nm}^{\rho} f_{nm}^{\sigma}+2 c_{nm}^{\sigma} f_{nm}^{\rho}
 - c_{nm}^{\sigma} f_{nm}^{\sigma})
\\ & &
 - \sum_{k} \alpha_{nm,k} (c_{nk}^{\rho}c_{km}^{\sigma}
  +c_{nk}^{\sigma}c_{km}^{\rho} + \frac{1}{2}c_{nk}^{\sigma} c_{km}^{\sigma})
 \end{eqnarray*}
  where $m\neq n$, and
 \begin{eqnarray*}
\dot{c}_{nn}^{\rho} & = &
 (c_{nn}^{\rho})^2+\frac{3}{16}(c_{nn}^{\sigma})^2)
\\ & &
 -\sum_{k}\alpha_{ii,k} (c_{nk}^{\rho}c_{kn}^{\rho}
  +\frac{3}{16}c_{nk}^{\sigma}c_{kn}^{\sigma})\\
\dot{c}_{nn}^{\sigma} & = &
 [2 c_{nn}^{\rho} c_{nn}^{\sigma}-\frac{1}{2}(c_{nn}^{\sigma})^2]
\\ & &
-\sum_{k}\alpha_{nn,k}(c_{nk}^{\rho}c_{kn}^{\sigma}
  +c_{nk}^{\sigma}c_{kn}^{\rho} + \frac{1}{2}c_{nk}^{\sigma} c_{kn}^{\sigma}).
\end{eqnarray*}
In the above
\begin{equation}
f_{nm}=\frac{ \tilde{f}_{nm}}{\pi(v_{Fn}+v_{Fm})},~~c_{nm}=\frac{ \tilde{c}_{nm}}{\pi(v_{Fn}+v_{Fm})},
\end{equation}
and
\begin{equation}
\alpha_{nm,k}=\frac{(v_{Fn}+v_{Fk})(v_{Fk}+v_{Fm})}{2v_{Fk}(v_{Fn}+v_{Fm})}.
\end{equation}
Moreover the dot in e.g., $\dot{f}_{nm}^{\rho}$ denotes the derivative with respect to
the logarithm of the energy cutoff.

The differences between the 2-leg pnictide ladder the the 5-leg one band Hubbard ladder lie in the values of the
Fermi velocities $v_{Fn}$ and the initial values of the coupling constants. Specifically the  initial coupling constants are determined by the parameters in Eq. \ref{model_5band} and the band wavefunctions as follows
\begin{widetext}
\begin{eqnarray*}
\tilde{f}_{nm}^{\rho} & = &
\frac{U_1}{2} \sum_{a}( u_{Rna} u_{Lma} u^*_{Lma} u^*_{Rna} )
+\frac{2U_2-J_H}{4} \sum_{a\neq b}[ u_{Rna} u_{Lmb}
 (u^*_{Lmb} u^*_{Rna}-\frac{1}{2} u^*_{Lma} u^*_{Rnb}) + (a \leftrightarrow b)]
\\ & &
+\frac{J_H}{2} \sum_{a\neq b}( u_{Rna} u_{Lma} u^*_{Rnb} u^*_{Lmb} ) \\
\tilde{f}_{nm}^{\sigma} & = &
 2 U_1 \sum_{a}( u_{Rna} u_{Lma} u^*_{Lma} u^*_{Rna} )
+ \sum_{a\neq b} [u_{Rna} u_{Lmb}
 (U_2 u^*_{Lma} u^*_{Rnb} - J_H u^*_{Lmb} u^*_{Rna})
  + (a \leftrightarrow b) ]
\\ & &
+J_H \sum_{a\neq b}( u_{Rna} u_{Lma} u^*_{Rnb} u^*_{Lmb} ) \\
\tilde{c}_{nm}^{\rho} & = &
 \frac{U_1}{2} \sum_{a} ( u_{Rna} u_{Lna} u^*_{Lma} u^*_{Rma} )
+\frac{2U_2-J_H}{4} \sum_{a\neq b}[ u_{Rna} u_{Lnb}
 (u^*_{Lmb} u^*_{Rma}-\frac{1}{2} u^*_{Lma} u^*_{Rmb}) + (a \leftrightarrow b)]
\\ & &
+\frac{J_H}{2} \sum_{a\neq b}( u_{Rna} u_{Lna} u^*_{Rmb} u^*_{Lmb} ) \\
\tilde{c}_{nm}^{\sigma} & = &
 2 U_1 \sum_{a}( u_{Rna} u_{Lna} u^*_{Lma} u^*_{Rma} )
+ \sum_{a\neq b} [u_{Rna} u_{Lnb}
 (U_2 u^*_{Lma} u^*_{Rmb} - J_H u^*_{Lmb} u^*_{Rma})
  + (a \leftrightarrow b) ]
\\ & &
+J_H \sum_{a\neq b}( u_{Rna} u_{Lna} u^*_{Rmb} u^*_{Lmb} ).
\end{eqnarray*}
\end{widetext}
In the above equation  $a,b=1,\dots,5$ label the five d-orbitals of Fe,
$u_{R/Lna}, a=1,...5 $ are the five orbital components of the Bloch wave functions associated with
the $R/L$ members of the $n$th Fermi pair.
Finally $(a\leftrightarrow b)$ denotes the exchange the subscript $a$ and $b$ in
the term right before it.

We stress here that most of the above coupling constants have the dual character
described in earlier sections.
For example, the term
 $\tilde{c}_{24}^{\rho} J_{24}^{R} J_{24}^{L} =
 \sum_{\alpha,\beta} \tilde{c}_{24}^{\rho}
 \psi_{R2\alpha}^\dagger \psi_{L2\beta}^\dagger
 \psi_{L4\beta}^{\vphantom{\dagger}}\psi_{R4\alpha}^{\vphantom{\dagger}}
$ can be viewed as both Cooper scattering and SDW scattering
between hole pocket 2 and electron pocket 4.  In this sense, the ladder RG overemphasis
these special channels. As a result the ladder RG retains the key ingredient which allow the SDW scattering to drive other ordering tendencies.

\begin{figure}[tbp]
\begin{center}
\includegraphics[angle=0,scale=0.43]
{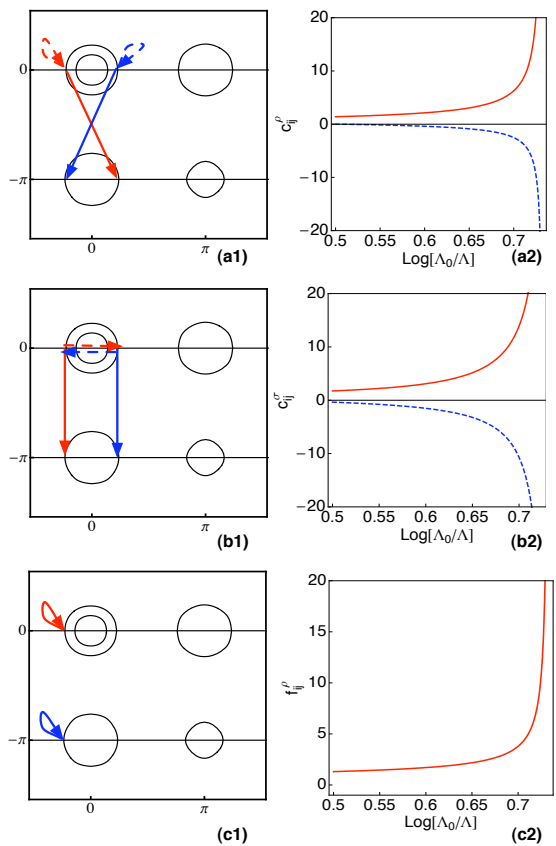}\caption{(color on-line) Left panels: the schematic representation of the strongest scattering processes (their scattering amplitude is at least two orders of magnitudes larger than all other scattering amplitudes) for the 2-leg pnictide ladder at low energies: (a) $c^\rho_{24}$ (solid line) and $c^\rho_{22}$ (dashed line) (b) $c^\sigma_{24}$ and $c^\sigma_{22}$ (dashed line) (c) $f^\rho_{24}$ (solid line). Right panels: the RG flow associated with the processes associated with the right panels.  \label{ladder_scattering}}
\end{center}
\end{figure}

Numerically we found that
all the coupling constants diverge at a finite cutoff.
Moreover the coupling constant associated with two Fermi pairs, one from the electron FS and the other from the hole FS, always dominate over others. For example, with the undoped band structure and initial couplings determined
by the same set of parameters, $U_1=4$, $U_2=2$, $J_H=0.7$, used in Sec.II we found that the coupling constants associated with Fermi pair $2$ (from the large hole pocket around $(0,0)$) and $4$ (from the electron pocket around $(0,-\pi)$) are larger than others  by more than two orders of magnitude.
More specifically the coupling constants with the largest magnitudes are
\be &&\tilde{c}_{24}^{\rho},~~\tilde{c}_{24}^{\sigma},~~\tilde{c}_{22}^{\rho},~~\tilde{c}_{22}^{\sigma},~~
 \tilde{c}_{44}^{\rho},~~\tilde{c}_{44}^{\sigma},~~\tilde{f}_{24}^{\rho}.\ee
The schematic representation of these scattering processes and their RG flow are plotted as Fig. \ref{ladder_scattering}.

As mentioned before $\tilde{c}_{24}^{\rho}$ has the dual character of SDW and Cooper scattering (solid line of Fig. \ref{ladder_scattering}(a)), it grows to large positive value hence drives SC pairing with reversed sign between electron and hole pocket.
$\tilde{c}_{24}^{\sigma}$ has the dual character of Cooper scattering and CDW (solid line of Fig. \ref{ladder_scattering}(b)), it also flows to large positive value, and contributes to both $s_{\pm}$ SC and imaginary CDW order. Here we note a little departure from the 2D physics in the sense that in 2D it is the dual SDW and CDW scattering that drives imaginary CDW. Here due to the omission of the Umklapp process $LL\rightarrow RR$ between $n=2$ and $n=4$ the process in Fig.7(a) is disallowed.
$\tilde{c}_{22}$ and $\tilde{c}_{44}$ are intra-pocket Cooper scattering (dashed line in Fig. \ref{ladder_scattering}(a,b)), they are driven by the inter-pocket Cooper scattering (which favors the $s_{\pm}$ pairing) and become large negative at the lowest energies. $\tilde{f}_{24}$ is the shared channel between SDW and forward scattering (solid line in Fig. \ref{ladder_scattering}(c)), it flows to large positive and causes PI as discussed in Sec II. It is interesting to observe that all important competing orders found
in 2D FRG are reproduced in this 1D ladder RG.

Pairing form factors are studied in the same way as in Sec.II. In this case \Eq{SCvertex}
are $5\times 5$ matrices ($n,m=1,\dots,5$)
\begin{eqnarray*}
V_{\rm SC,s}(n,m,\Lambda) & = & \tilde{c}_{nm}^{\rho}
 +\frac{3}{4}\tilde{c}_{nm}^{\sigma},\\
V_{\rm SC,t}(n,m,\Lambda) & = & \tilde{c}_{nm}^{\rho}
 -\frac{1}{4}\tilde{c}_{nm}^{\sigma}
\end{eqnarray*}
and negative eigenvalues $w_i$ indicate pairing instability.
We found that the triplet channels are much weaker than the singlet ones,
and among the five singlet channels only one of them become negative upon
RG flow Fig. \ref{ladderRG}(b)
Eigenvector(form factor) corresponding to this channel has
opposite signs between the hole pocket 2 and electron pocket 4,
consistent with the $s_{\pm}$ pairing found in 2D FRG,
but has nearly vanishing weights on the other Fermi points.

\section{Conclusion}

Motivated by the high T$_c$ superconductivity of iron pnictide, we study the pure electronic pairing mechanism in both the cuprates and the pnictides from an itinerant point of view, where the FRG method becomes very useful. The main message of this paper is a view point we obtained from the FRG studies,  namely, in a  system with strong antiferromagnetic correlation such as the cuprate and the iron pnictide, there are generically SC, PI and orbital-current (CDW) ordering tendencies. These ordering tendencies are triggered by a set of dual-character scattering processes associated with SDW-SC, SDW-PI and SDW-CDW, From these dual processes, other generic SC, PI and CDW scattering grow. Thus it is through these dual processes that SDW fluctuation drives all the other instabilities. Since these SDW related dual processes all have positive amplitudes, it imposes strong constraint on the symmetry of the form factor associated with the induced order parameters. The details as summarized in the table below.

Other significant results presented in this paper are also summarized as follows.  (i) The origin of the SC gap anisotropy in iron pnictides. We argue the SC gap anisotropy in the iron pnictide is due to the variation of the orbital content of the Bloch wave function along the FS. (ii) We show that the Pomerachuk instability in the iron pnictide tends to shrink   both the electron and hole pockets. It can give rise to a change of the FS topology of the electron pockets. (iii) We demonstrate the validity of the $J_1-J_2$ model in describing the lowest energy effective Hamiltonian for the iron pnictides, and determine the range of $J_1-J_2$ for which this model is valid. We commented on the inadequacy of
the $J_1-J_2$ model in the intermediate energy regime where competing order is manifested. (iv) We discuss the ordering tendencies of the cuprates including charge stripes under the same framework.  (v) We show that the double-layer Hubbard model nicely interpolates between the physics of the cuprates and the physics of the pnictides as a function of the interlayer hopping parameter. We predict the existence of a quantum phase transition where the superconducting pairing changes from $d_{x^2-y^2}$-wave to the $s_{\pm}$ symmetry.  (vi) We present the renormalization group results for a  2-leg ladder version of the iron pnictides Hamiltonian. We show that the results are qualitatively consistent with those obtained from two-dimensional functional renormalization group calculations.

\acknowledgments
DHL was supported by DOE grant number DE-AC02-05CH11231.

\begin{widetext}
\begin{center}
\begin{table}[htdp]
\begin{tabular}{|c|c|c|c|c|c|}
\hline
      & Dual processes  &  Two ways of decoupling          &     Implication      &         Cuprate        &       Iron Pnictide               \\
 \hline
SDW/SC  &  $Vc^\dag_{{\bf k+Q}s}c^\dag_{{\bf -k+Q}s^\prime}c_{{\bf k}s^\prime}c_{{\bf -k}s}$ & $-V\vec{S}_{{\bf k}}\vec{S}_{{\bf -k}}$ or $V\Delta^\dag_{{\bf k+Q}}\Delta_{{\bf k}}$ &  $\langle \Delta_{{\bf k}}\rangle\langle \Delta_{{\bf k+Q}}\rangle<0$ &${\bf Q}=(\pi,\pi)$,   &   ${\bf Q}=(\pi,0)$/$(0,\pi)$, \\
  &                  $V>0$                            &                              $  \Delta^\dag_{{\bf k}}=c^\dag_{{\bf k}s}c^\dag_{{\bf -k}s^\prime}$           &  (*)                 &both $\cos k_x+\cos k_y$     & both  $\cos k_x\cos k_y$  \\
    &                                            &          $\vec{S}_{{\bf k}}=\sum\limits_{ss^\prime} c^\dag_{{\bf k+Q}s}\vec{\sigma}_{ss^\prime}c_{{\bf k}s^\prime}$                                        &                     &  and $\cos k_x-\cos k_y$   &  and $\sin k_x\sin k_y$  \\
        &                                             &                                                &                     & satisfy (*) &  satisfy (*)  \\
     &                                             &                                               &                     & FS determines & FS determines  \\
       &                                             &                                                 &                     & $\cos k_x-\cos k_y$ &  $\cos k_x\cos k_y$  \\
  \hline
  SDW/PI  &  $Vc^\dag_{{\bf k+Q}s}c^\dag_{{\bf k}s^\prime}c_{{\bf k}s^\prime}c_{{\bf k+Q}s}$ & $-V\vec{S}_{{\bf k}}\vec{S}_{{\bf k+Q}}$ or $Vn_{{\bf k+Q}}n_{{\bf k}}$ & $\delta n_{\v k}\delta n_{\v k+\v Q}<0$ & $C_{4v}$ breaking   & Shrinking of all \\
   &     $V>0$                                        &                                 $n_{{\bf k}}=\sum\limits_s c^\dag_{{\bf k}s}c_{{\bf k}s}$ &         &  FS distortion  & pockets  \\
      &                                             &             $\vec{S}_{{\bf k}}=\sum\limits_{ss^\prime} c^\dag_{{\bf k+Q}s}\vec{\sigma}_{ss^\prime}c_{{\bf k}s^\prime}$                                       &          &  See Fig. \ref{oneband_orders}(c)  &    See Fig. \ref{PI}(c,d) \\
      &                                             &                                                  &          &  & \\

\hline
SDW/CDW  &  $Vc^\dag_{{\bf k+Q}s}c^\dag_{{\bf k+Q}s^\prime}c_{{\bf k}s^\prime}c_{{\bf k}s}$ & $-V\vec{S}_{{\bf k}}\vec{S}_{{\bf k}}$ or $Vd_{{\bf k}}d_{{\bf k}}$ & $\langle d_{\v k}\rangle=$imaginary  & DDW & $(\pi,0)$/$(0,\pi)$  \\
  &     $V>0$                                        &                               $d_{{\bf k}}=\sum\limits_s c^\dag_{{\bf k+Q}s}c_{{\bf k}s}$                   &   (orbital current)                      &   See Fig. \ref{oneband_orders}(d) & orbital current order    \\
   &                                             &               $\vec{S}_{{\bf k}}=\sum\limits_{ss^\prime} c^\dag_{{\bf k+Q}s}\vec{\sigma}_{ss^\prime}c_{{\bf k}s^\prime}$                                    &         &    &  See Fig. \ref{CDW}\\

\hline
\end{tabular}
\caption{Summary of spin density wave fluctuation driven superconducting pairing, Pomeranchuk instability and charge current density wave \label{tab}}
\end{table}
\end{center}
\end{widetext}

\end{document}